\documentclass[twocolumn,tighten,times]{aastex631}

\usepackage{amsmath}
\usepackage{bm}
\usepackage{soul}
\DeclareMathOperator{\Epps}{Epps}
\shorttitle{SMBH Growth in Ellipticals}
\shortauthors{Farrah, Petty, Croker, et al.}
\newcommand{\uhm}{Department of Physics and Astronomy, University of Hawai`i at M\=anoa, 2505 Correa Rd., Honolulu, HI, 96822, USA}

\begin{document}

\title{A Preferential Growth Channel for Supermassive Black Holes in Elliptical Galaxies at $z\lesssim2$}

\correspondingauthor{Duncan~Farrah}
\email{dfarrah@hawaii.edu}

\author[0000-0003-1748-2010]{Duncan~Farrah}
\affiliation{\uhm}
\affiliation{Institute for Astronomy, University of Hawai`i,  2680 Woodlawn Dr., Honolulu, HI, 96822, USA}
\author[0000-0003-0624-3276]{Sara~Petty}
\affiliation{NorthWest Research Associates, 3380 Mitchell Ln., Boulder, CO 80301, USA}
\affiliation{Convent \& Stuart Hall Schools of the Sacred Heart, 2222 Broadway, San Francisco, CA 94115, USA}
\author[0000-0002-6917-0214]{Kevin~S.~Croker}
\affiliation{\uhm}
\author[0000-0003-1704-0781]{Gregory~Tarl\'e}
\affiliation{Department of Physics, University of Michigan, 450 Church St., Ann Arbor, MI, 48109, USA}
\author[0000-0002-0147-0835]{Michael~Zevin}
\affiliation{Kavli Institute for Cosmological Physics, The University of Chicago, 5640 South Ellis Avenue, Chicago, IL 60637, USA}
\affiliation{Enrico Fermi Institute, The University of Chicago, 933 East 56th Street, Chicago, IL 60637, USA}
\author[0000-0003-0917-9636]{Evanthia Hatziminaoglou}
\affiliation{ESO, Karl-Schwarzschild-Str 2, D-85748 Garching bei München, Germany}
\author[0000-0001-8973-5051]{Francesco Shankar}
\affiliation{Department of Physics and Astronomy, University of Southampton, Highfield SO17 1BJ, UK}
\author[0000-0002-6736-9158]{Lingyu Wang}
\affiliation{Kapteyn Astronomical Institute, University of Groningen, Postbus 800, 9700 AV Groningen, the Netherlands}
\affiliation{SRON Netherlands Institute for Space Research, Landleven 12, 9747 AD, Groningen, the Netherlands}
\author[0000-0002-9548-5033]{David L Clements}
\affiliation{Imperial College London, Blackett Laboratory, Prince Consort Road, London, SW7 2AZ, UK}
\author[0000-0002-2612-4840]{Andreas Efstathiou}
\affiliation{School of Sciences, European University Cyprus, Diogenes Street, Engomi, 1516 Nicosia, Cyprus}
\author[0000-0002-3032-1783]{Mark Lacy}
\affiliation{National Radio Astronomy Observatory, Charlottesville, VA, USA}
\author[0000-0001-8818-8922]{Kurtis~A.~Nishimura}
\affiliation{\uhm}
\author[0000-0002-9149-2973]{Jose Afonso}
\affiliation{Instituto de Astrof\'{i}sica e Ci\^{e}ncias do Espa\c co, Universidade de Lisboa, Portugal}
\affiliation{Departamento de F\'{i}sica, Faculdade de Ci\^{e}ncias, Universidade de Lisboa, Portugal}
\author[0000-0001-6139-649X]{Chris Pearson}
\affiliation{RAL Space, STFC Rutherford Appleton Laboratory, Didcot, Oxfordshire OX11 0QX, UK}
\affiliation{The Open University, Milton Keynes MK7 6AA, UK}
\affiliation{Oxford Astrophysics, University of Oxford, Keble Rd, Oxford OX1 3RH, UK}
\author[0000-0002-5206-5880]{Lura K Pitchford}
\affiliation{Department of Physics and Astronomy, Texas A\&M University, College Station, TX, USA} 
\affiliation{George P. and Cynthia Woods Mitchell Institute for Fundamental Physics and Astronomy, Texas A\&M University, College Station, TX, USA}

\begin{abstract}
The assembly of stellar and supermassive black hole (SMBH) mass in elliptical galaxies since $z\sim1$ can help to diagnose the origins of locally-observed correlations between SMBH mass and stellar mass. We therefore construct three samples of elliptical galaxies, one at $z\sim0$ and two at $0.7\lesssim z \lesssim2.5$, and quantify their relative positions in the $M_{BH}-M_*$ plane. Using a Bayesian analysis framework, we find evidence for translational offsets in both stellar mass and SMBH mass between the local sample and both higher redshift samples. The offsets in stellar mass are small, and consistent with measurement bias, but the offsets in SMBH mass are much larger, reaching a factor of seven between  $z\sim1$ and $z\sim0$. The magnitude of the SMBH offset may also depend on redshift, reaching a factor of $\sim20$ at $z\sim 2$. The result is robust against variation in the high and low redshift samples and changes in the analysis approach. The magnitude and redshift evolution of the offset are challenging to explain in terms of selection and measurement biases. We conclude that either there is a physical mechanism that preferentially grows SMBHs in elliptical galaxies at $z\lesssim 2$, or that selection and measurement biases are both underestimated, and depend on redshift. 
\end{abstract}


\section{Introduction}
At low redshift, more massive black holes tend to reside in more massive galaxies. This tendency can be parameterized by the ratio of supermassive black hole (SMBH) mass to host stellar mass --- $M_{BH}/M_{*}$ --- and has been considered across a wide range in galaxy stellar mass \citep[e.g.][]{mag98,gebh00,marconi03,feme04,haring04,kormendyho13,mccma13,schutte19,zhao21}. Several different physical origins have been proposed for this correlation. At least in part, it may arise because both star formation and SMBH accretion are fueled by a common gas reservoir \citep[e.g.][]{peng07,jahnke11}. It may also arise because star formation and SMBH accretion can trigger and/or quench each other, in a family of processes termed `feedback' \citep[e.g.][]{fabian12,farrah12,gonza17}.

The redshift evolution of the $M_{BH}/M_{*}$ ratio may give insight into the processes that shape the SMBH--stellar mass correlation. Studies of the redshift evolution of the $M_{BH}/M_{*}$ ratio have focused on active galactic nuclei (AGN), since SMBH masses at $z\gtrsim0.1$ can currently only be estimated in AGN. These studies have found conflicting results. Some find evidence for positive evolution in the $M_{BH}/M_{*}$ ratio with redshift --- that is, relatively more massive SMBHs at higher redshifts \citep{decarli10,merloni10,benn11,ding20}. Others find negative evolution \citep{ueda18}, or no evolution \citep{shields03,cist11,schramm13,sun15,suh20,lijen21,lisync21}. Potential reasons discussed by these authors for observed redshift evolution (or lack thereof) in the $M_{BH}/M_{*}$ ratio include an evolving (or constant) dark matter halo mass threshold for AGN activity, AGN feedback, or a combination of selection and measurement bias in SMBH and stellar mass measurements. A consensus, however, has yet to emerge.

Complementary insight can be gained by constructing an experiment that differs in two ways from previous studies. First, is to use alternate diagnostics to the redshift evolution of the $M_{BH}/M_{*}$ ratio. One such diagnostic are the positions of galaxies along the $M_{BH}$ and $M_{*}$ axes in the $M_{BH} - M_{*}$ plane. Treating these positions as independent degrees of freedom may encode more information on channels by which SMBH and stellar mass can assemble. Second is to restrict consideration to specific galaxy types. Because early- and late-type galaxies have distinct evolutionary paths, restriction of galaxy type can reduce the diversity of processes that shape assembly, thus simplifying interpretation of results. In this context, the simplest galaxies are massive ellipticals, due to their quiescence and lack of recent major assembly activity.

In this paper, we select three samples of ellipticals; one that represents the quiescent population of local ellipticals, and two at $z>0.8$ that represent the emergence of ellipticals onto the red sequence. The expectation for these galaxies  is insignificant change in $M_*$ and $M_{BH}$ with redshift. We compare the high-redshift samples against the low-redshift sample to test for stellar and/or SMBH mass assembly between them. Any changes in position within the $M_{BH} - M_{*}$ plane can then be used to diagnose how passively evolving elliptical galaxies assemble stellar and/or SMBH mass.

The structure of this paper is as follows. In \S\ref{sec:sample_selection} we introduce the catalogs used in our study and describe how we select systems to plausibly enforce a passive ancestral relation. In \S\ref{sec:model_parameters} we describe the Bayesian framework we use to compare the high and low redshift samples. In \S\ref{sec:results} we present our results. In \S\ref{sec:discussion}, we discuss systematic and physical origins of our results. Throughout, we assume \mbox{$H_0 = 70$\,km\,s$^{-1}$\,Mpc$^{-1}$}, \mbox{$\Omega = 1$}, and \mbox{$\Omega_{\Lambda} = 0.7$}. We convert all literature data to this cosmology where necessary. We use the word `draw' instead of `sample' to distinguish between the selected samplings of high and low redshift populations and realizations drawn from the distributions in our analysis. Overset tilde will denote median values.

\begin{figure}
\begin{center}
\includegraphics[width=0.98\linewidth]{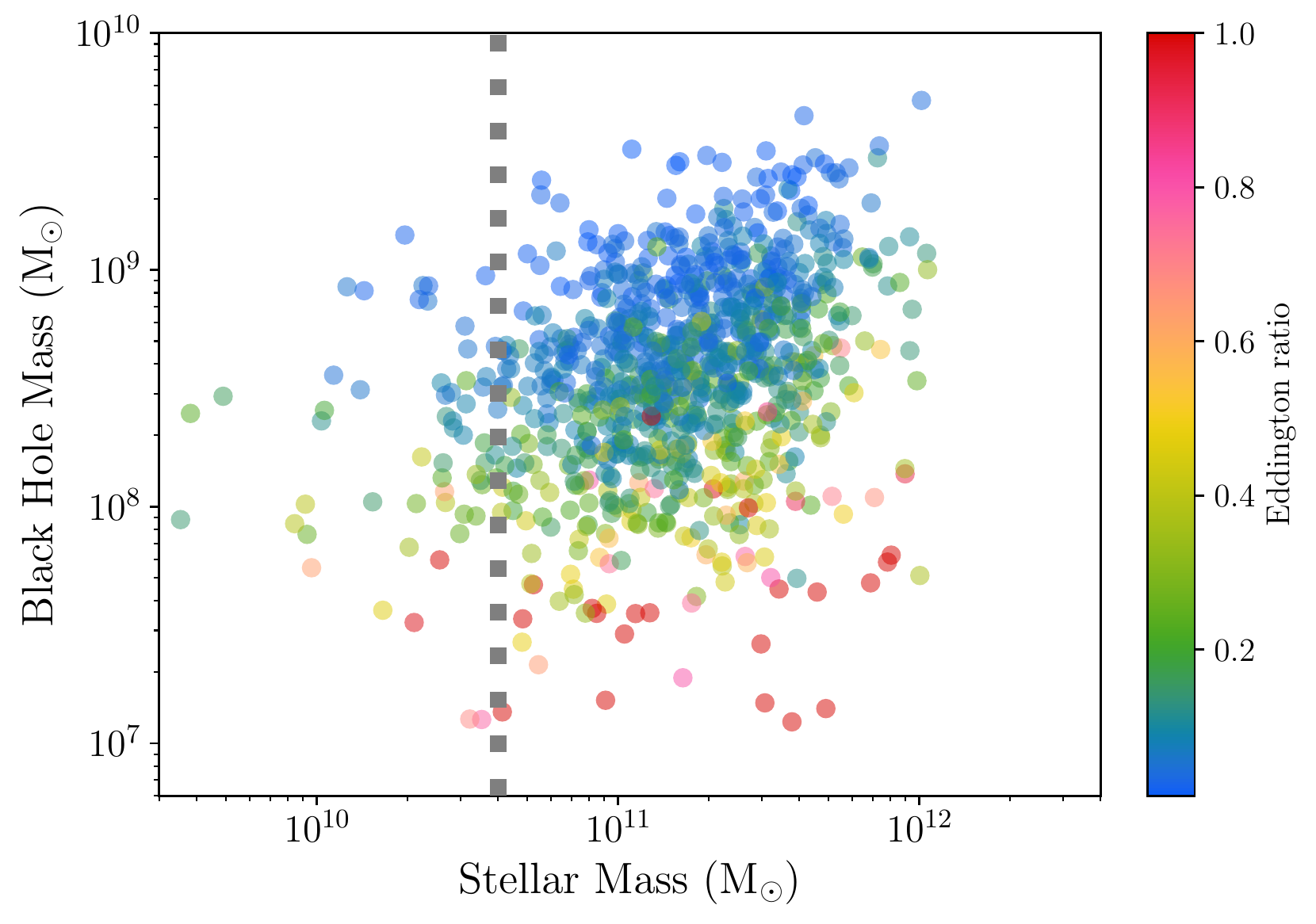} 
\includegraphics[width=0.98\linewidth]{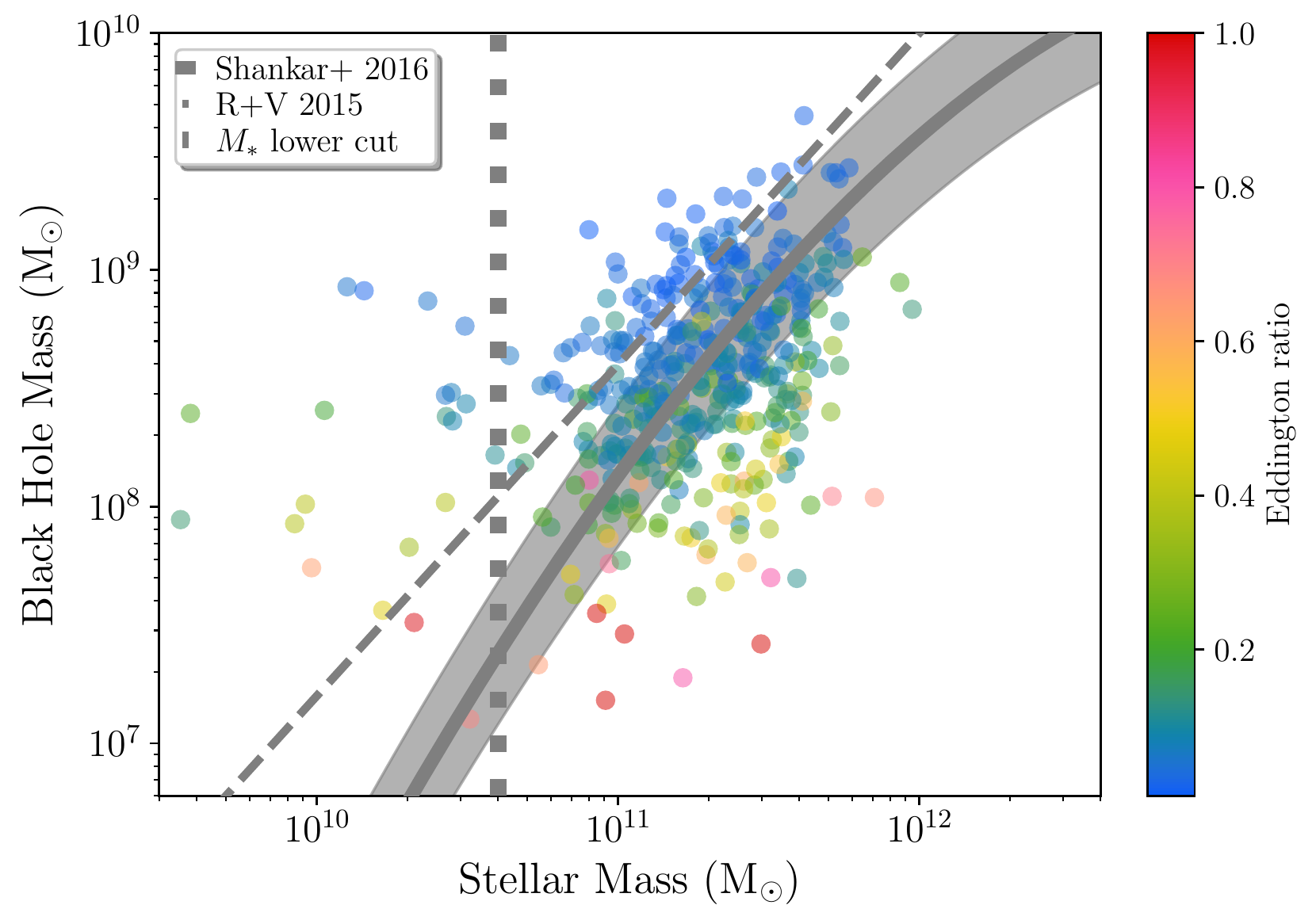} 
\caption{Diagnostics of our high redshift WISE sample (\S\ref{sec:hizwise}). {\itshape Top:} The distributions of all AGN in \citet{barrows21} at $0.8<z<0.9$ with Elliptical host SEDs in the $M_{BH} - M_{*}$ plane, before our SFR and $E_{B-V}$ cuts are applied. We overplot our $M_{*}$ selection boundary. {\itshape Bottom:} Same as the middle panel, but with our SFR and $E_{B-V}$ cuts applied. We overplot two literature relations. First, the \citet{rein15} $M_{BH} - M_{*}$ relation for quiescent ellipticals (see also e.g. \citealt{savgra16,bentz18,benn21}). Second, a proposed intrinsic $M_{BH} - M_{*}$ relation (\citealt{shank16a}, \S\ref{sec:prior_Bdyn}).}
\label{fig:samplewise}
\end{center}
\end{figure}

\section{Sample Selection}\label{sec:sample_selection}
Our sample selection is divided into high redshift (\S\ref{sec:hiz}) and low redshift (\S\ref{sec:local}) samples. At high redshifts, only single-epoch virial SMBH masses are widely available, so we select AGN in elliptical hosts that plausibly represent the final unobscured AGN phase in an emerging elliptical. At low redshift, we select Ellipticals based on the availability of stellar dynamical SMBH masses.

\subsection{High redshift samples}\label{sec:hiz}

\subsubsection{WISE}\label{sec:hizwise}
Assembling large samples of high-redshift AGN in elliptical hosts is possible via the catalog presented by \citet{barrows21}, which is based on observations with the Wide-field Infrared Survey Explorer (WISE, \citealt{wright10}). The \citet{barrows21} catalog comprises 695,273 AGN jointly selected from the WISE, Galaxy Evolution Explorer (GALEX), and Sloan Digital Sky Survey (SDSS) DR14 surveys. Host properties were determined using CIGALE \citep{boqu19}, assuming a Salpeter \citep{salpeter55} initial mass function (IMF). The spectral energy distribution (SED) fits incorporate near- and mid-infrared data, where the host-AGN contrast is low, thus minimizing bias towards more massive hosts for a given AGN luminosity. The key catalog parameters are host type (elliptical, spiral, or irregular), host star formation rate (SFR), and AGN reddening ($E_{b-v}$). We obtained SMBH masses for the AGN by cross-matching with the catalog of optically selected, broad-line AGN in \citet{raks20}, using a $2\arcsec$ radius. Their SMBH masses are computed using the prescriptions of \citet{vestpe06} for H$\beta$ and \citet{vestos09} for \ion{Mg}{2}. This results in 160,004 matches. Because some later comparisons assume a Chabrier IMF \citep{chabrier03}, we convert their stellar masses to a Chabrier IMF, multiplying by $0.61$: the ratio of these two integrated IMFs.

To select unobscured AGN in elliptical hosts, we apply the following criteria:

\begin{enumerate}

\item A redshift range of $0.8 < z < 0.9$. This is within the epoch of emergence of the red sequence, and ensures that all the SMBH masses are measured using the \ion{Mg}{2} line.
    
\item Minimally reddened AGN, with $E_{B-V}<0.2$.
    
\item An elliptical host SED, with contributions from other host types of $\leqslant5$\% of the elliptical value. Allowing for a small contribution from other host SED types enables selection of elliptical hosts with a UV excess. There is no consistent morphological information available for the WISE sample, so we rely on their host SED classifications.

\item A host SFR  at least a factor of five below the SFR--$M_{*}$ `main sequence' at the redshift of the source \citep[equation 26 of][]{speagle14}. This selects for quiescent hosts while factoring in the redshift evolution of quiescence.

\end{enumerate}

This results in a sample of 420 objects. We plot diagnostics of this sample in Figure \ref{fig:samplewise}.

\begin{figure}
\begin{center}
\includegraphics[width=0.98\linewidth]{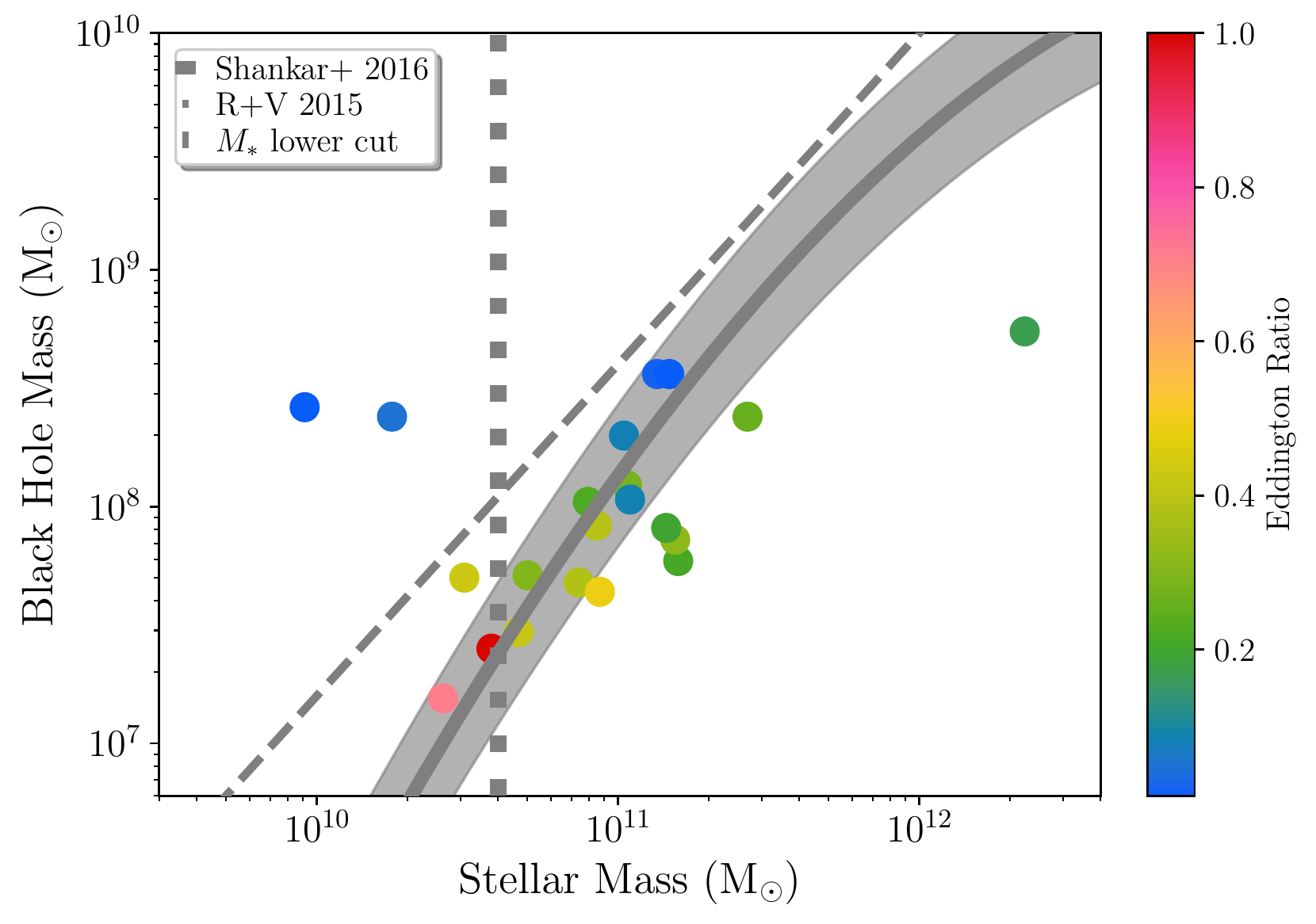} 
\caption{The distribution of the COSMOS sample in the $M_{BH} - M_{*}$ plane (\S\ref{samp:cosmosm}). Also plotted are the $M_{*}$ selection boundary, the \citet{rein15} $M_{BH} - M_{*}$ relation for local quiescent ellipticals, and the proposed intrinsic $M_{BH} - M_{*}$ relation from \citet{shank16a}.}
\label{fig:samplecosmos}
\end{center}
\end{figure}

\subsubsection{COSMOS}\label{samp:cosmosm}
To extend our analysis to higher redshifts, we start with the sample in the COSMOS field from \citet{suh20}. This sample includes 100 broad-line AGN, and is drawn from a parent catalog of $\sim4000$ X-ray sources with sensitivity to SMBH mass down to $\sim10^{7}$\,M$_{\odot}$ at $z=2$. This sample complements the WISE sample in two ways. First, the deeper COSMOS observations enable finding quiescent hosts further into the epoch of formation of the red sequence. Second, the X-ray rather than optical selection, and a different stellar mass calculation approach, offer a cross-check on the WISE results. The SMBH masses are calculated using the calibration of \citet{vestpe06} for H$\alpha$ and H$\beta$ and \citet{trakh12} for \ion{Mg}{2}. The assumed virial factor is thus $f_{vir}\sim 4-5$. The prescriptions used to calculate SMBH masses by \citet{suh20} and \citet{raks20} are identical for H$\beta$, but differ for \ion{Mg}{2}. We thus recompute the \ion{Mg}{2}-based SMBH masses using the prescription used by \citet{raks20}. This corresponds to a downward correction by $\sim0.17$\,dex. The stellar masses are derived via template fitting \citep{suh19} and assume a Chabrier IMF.

To assemble ancillary data that aid in selecting for quiescent hosts, such as SFRs, we cross-match this sample with the data in \citet{bong12}. To select quiescent hosts in the COSMOS sample, we adopt the same selection boundaries on AGN reddening ($E_{B-V}<0.1$) and position below the SFR -- $M_{*}$ main sequence at the redshift of the source. We adopt a redshift range of $0.7 < z < 2.5$, which gives overlap with the WISE sample.

It is not possible to perform a selection on host SED type for the COSMOS sample. Instead, we use the Hubble Space Telescope (HST) imaging in the COSMOS field to perform a basic selection on morphology. To do so, we first band-merge our catalog with the morphological catalog of AGN in the COSMOS field presented by \citet{griffith10}. Even with HST data though, morphological classification at these redshifts is challenging against AGN glare. We therefore perform only basic selections. We exclude objects classified as either unresolved point sources or disks. We do not exclude objects with an `unknown' morphology. This gives a sample of 23 objects. We plot this sample in the $M_{BH} - M_{*}$ plane in Figure \ref{fig:samplecosmos}.

\begin{figure}
\begin{center}
\includegraphics[width=\columnwidth]{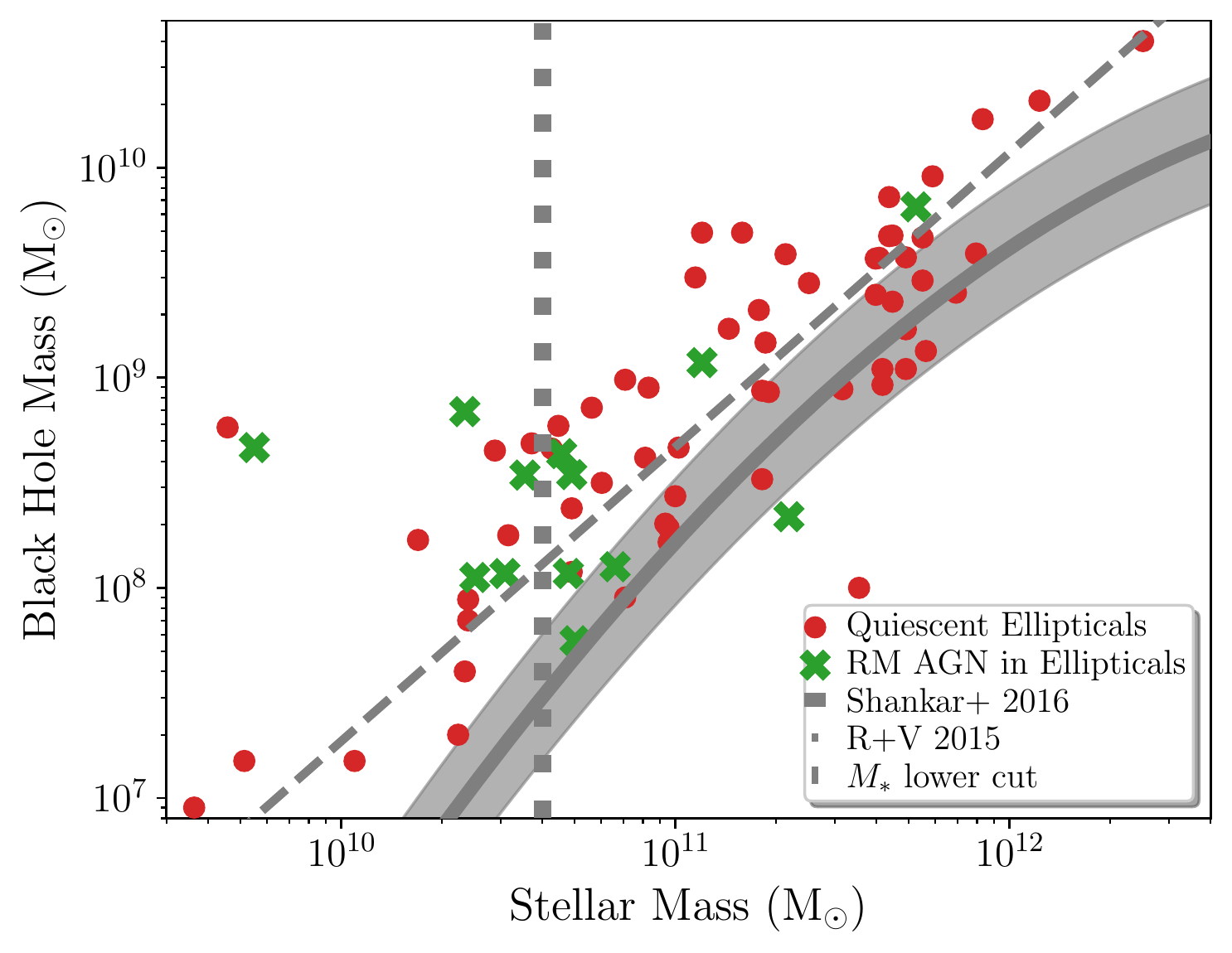}
\caption{The distributions of the low-redshift quiescent sample (\S\ref{sec:quiesc}) and RM AGN (\S\ref{sec:rmagn}) sample in the $M_{BH} - M_{*}$ plane. The distributions of the two samples are similar. Also plotted are the $M_{*}$ selection boundary, the \citet{rein15} $M_{BH} - M_{*}$ relation for local quiescent ellipticals, and the proposed intrinsic $M_{BH} - M_{*}$ relation from \citet{shank16a}.}
\label{fig:samplelocal}
\end{center}
\end{figure}

\subsection{Low-redshift Samples}\label{sec:local}

\subsubsection{Quiescent Sample}\label{sec:quiesc}
The criteria for the low-redshift sample are that they be early-type quiescent galaxies with measured SMBH and stellar masses, and no evidence for recent major assembly activity. To select this sample, we start with the compilation of \citet{zhu21}, which is based on the sample of \citet{kormendyho13}. The SMBH masses are calculated via stellar dynamical modeling, while the stellar masses are calculated from a multi-component fit to the 2MASS $K_{s}$-band imaging, via a mass-to-light (M/L) ratio from \citet{into13}, with a \citet{kroupa03} IMF. We note, but do not attempt to correct for, the possibility that stellar masses of galaxies can be underestimated when using 2MASS imaging \citep{laesker14}.  We further include the samples presented by \citet{thaler2019,yildirim16,yildirim17,yildirim17b,sahu19}, and \citet{pilawa22}. We convert to a Kroupa IMF where necessary. From this parent compilation, we first restrict to systems with a morphological classification as an elliptical or S0. We then exclude any system with a luminous AGN, a bar or a pseudobulge (as such non-axisymmetric features may prefer nuclear assembly, \citealt{hu08,gonza21}), and any evidence for a recent merger. This results in a sample of 86 objects, listed in Table~\ref{tbl:lowz_sample} and plotted in Figure \ref{fig:samplelocal}. 

The final sample has two issues. First, it is not homogeneously selected, as no such sample of quiescent ellipticals with measured SMBH masses currently exists. Requiring an SMBH mass from stellar dynamical modeling may also bias the sample to more massive SMBHs (\S\ref{sec:prior_Bdyn}). Second, the low-redshift sample is not explicitly matched in environment to the high-redshift sample. This however should not affect our results. High-redshift AGN reside, on average, in overdense regions \citep{croom05,myers06,ross2009,eftek17}, while our low-redshift sample spans environments ranging from the field to rich cluster cores (Table~\ref{tbl:lowz_sample}).

\subsubsection{Reverberation-mapped AGN}\label{sec:rmagn}
As a check on the quiescent sample, we assemble a sample of low-redshift AGN in elliptical hosts, selected to be plausibly brief `flare up' episodes within the quiescent elliptical population. This check is most readily performed with reverberation mapped (RM) BH mass measures because RM measured objects are thoroughly studied within the literature, which facilitates determination of host morphologies. 

We use published samples of RM AGN in early-type hosts \citep{bentz18,lisha21,hu21}, adopting \citet{into13} M/L ratio stellar masses to match that used by \citet{zhu21}. \citet{rein15} also present a RM AGN sample, but we cannot include it as none reside within confirmed elliptical hosts. Finally, we exclude any RM objects with signs of an ongoing merger. The RM AGN are not subject to the issues with completeness when $M_* \leqslant 4\times 10^{10} M_\odot$, so we improve statistics in this sample by adopting a lower $M_{*}$ cut of $10^{10}~M_\odot$. The sample is presented in Table~\ref{tbl:lowz_sample}, where we have also included the EHT SMBH mass measure for M87 \citep{ehtm8719}. We display the RM AGN sample in the $M_{BH} - M_{*}$ plane in Figure \ref{fig:samplelocal}. We discuss some aspects of the selection of this sample in \S\ref{biases:revagn}.

\section{Methods}\label{sec:model_parameters}
We aim to test the hypothesis that passively evolving ellipticals show no assembly of either $M_*$ or $M_{BH}$. To do so, we require an analysis pipeline that allows for the possibility of assembly in either $M_{BH}$ or $M_{*}$. In order to simultaneously consider possible mass assembly alongside measurement error, we perform a Bayesian analysis with the nested sampler \texttt{DYNESTY} \citep{skilling2004nested, feroz2009multinest, speagle2020dynesty}. These samplers measure Bayesian evidence, allowing quantitative comparisons of different models, and produce draws from posterior distributions. Our analysis requires defining three components; model parameters, prior distributions for these parameters, and a likelihood function.

The true distribution of objects in the $M_{BH} - M_{*}$ plane at any redshift is unknown. However, testing our hypothesis only requires examining whether this unknown distribution changes with redshift. Therefore, our model parameters do not describe the underlying distribution of the high and low redshift data $\left\{\mathbf{d}_h,\mathbf{d}_\ell\right\}$ within the $M_{BH}-M_*$ plane. Instead, our model parameters are:

\begin{itemize}
\item $\tau_{*}$: The translational offset between the high and low redshift samples along the stellar mass axis.
\item $\tau_{BH}$: The translational offset between the high and low redshift samples along the SMBH mass axis.	
\end{itemize}

In both cases, the translational offsets are defined as being applied to the high redshift sample.

For both parameters $\tau_{*}$ and $\tau_{BH}$ we adopt uniform and wide priors:

\begin{align}\label{eqn:tau_BH_decomposed}
 -1.0 \leqslant \tau_* \leqslant 1.0~&\mathrm{dex}   \\
 -0.5 \leqslant \tau_{BH} \leqslant 2.0~&\mathrm{dex}. 
\end{align}
in which both translations are given as offsets in log space.

Our likelihood function answers the question: what is the probability that the low-redshift sample and the $\bm{\tau}$-translated high-redshift sample are drawn from the same underlying distribution? Thus, if the data, as reported, are 1) unbiased, and 2) the high and low redshift populations are ancestral and passively evolving, then the analysis will recover median $\tau_* = \tau_{BH} = 0$. The details of our likelihood function, as well as other aspects of the Bayesian analysis, are presented in Appendix~\ref{sec:pipeline}.

\section{Results} \label{sec:results}
\begin{figure}
\begin{center}
\includegraphics[width=\columnwidth]{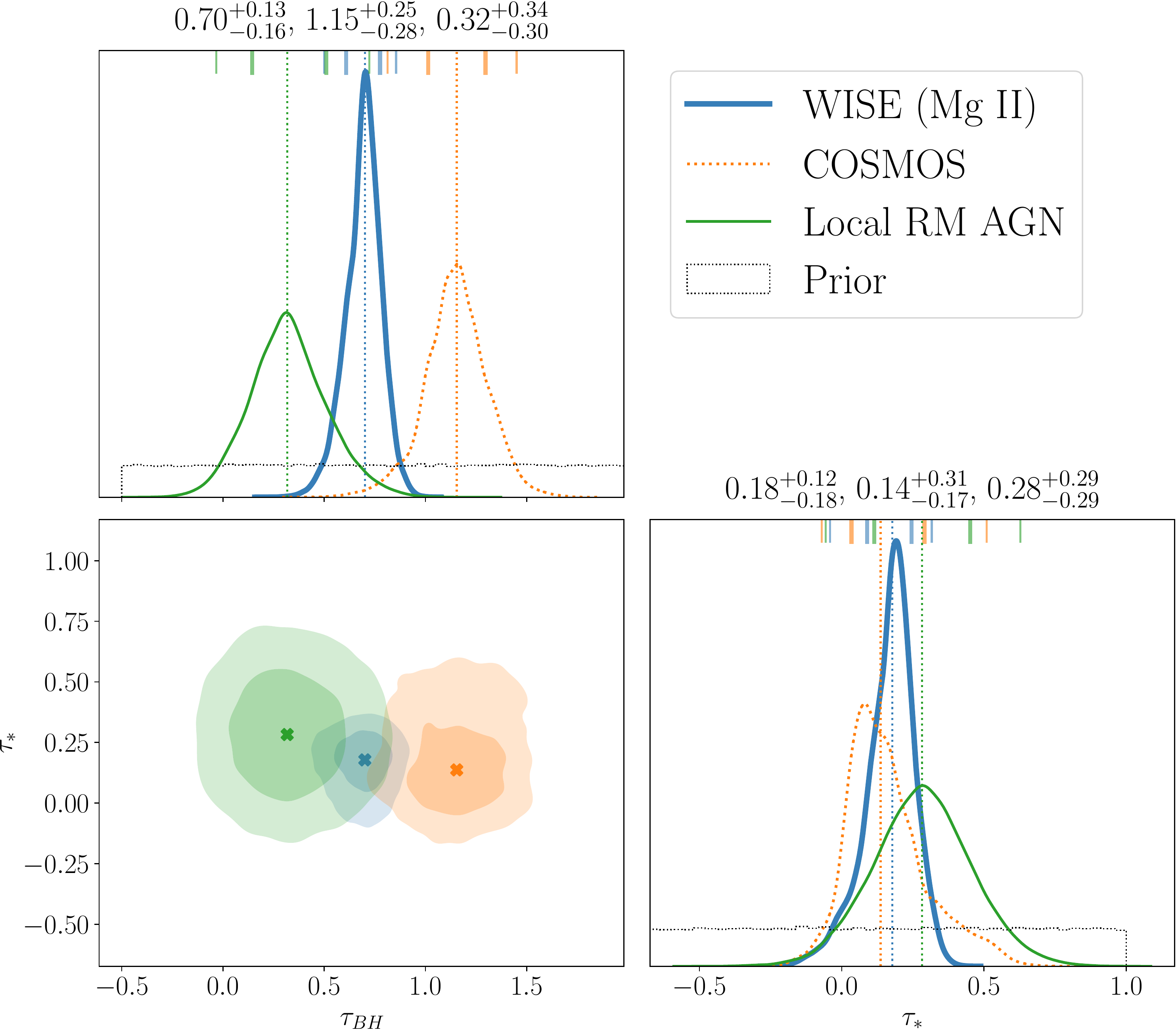}
\caption{\label{fig:anomaly}Posterior distributions for  $\tau_{BH}$ and $\tau_*$. These are the translations within the $M_{BH}-M_*$ plane required to align the high redshift WISE (blue), COSMOS (orange, dotted), and low redshift Local RM AGN (green, solid) samples with the low-redshift quiescent sample. Uncertainties give $90\%$ confidence. The WISE sample is at median $\widetilde{z} = 0.85$, the COSMOS sample is at median $\widetilde{z} = 1.61$, and the RM AGN sample is at median $\widetilde{z} = 0.02$.}
\end{center}
\end{figure}

The marginalized and joint posterior distributions for $\tau_{BH}$ and $\tau_*$ are reported in Figure~\ref{fig:anomaly}. The recovered translational offsets between the WISE sample and the low redshift quiescent sample are:

\begin{align}\label{eq:medtau_wise}
\tau_{BH} & = 0.70~^{+0.13}_{-0.16}~\mathrm{dex} \\
\tau_* & = 0.18~^{+0.12}_{-0.18}~\mathrm{dex},
\end{align}

\noindent and those for the COSMOS sample and the low redshift quiescent sample are:

\begin{align}
\tau_{BH} & = 1.15~^{+0.25}_{-0.28}~\mathrm{dex} \\
\tau_* & = 0.14~^{+0.31}_{-0.17}~\mathrm{dex}, \label{eq:medtaustar_cosmos}
\end{align}

\noindent For the RM AGN sample and the low redshift quiescent sample they are:
\begin{align}
\tau_{BH} & = 0.32~^{+0.34}_{-0.30}~\mathrm{dex} \\
\tau_* & = 0.28~^{+0.29}_{-0.29}~\mathrm{dex}, \label{eq:medtaustar_reverbs}
\end{align}
all at $90\%$ confidence.

\begin{figure*}
\begin{center}
 \includegraphics[width=0.49\linewidth]{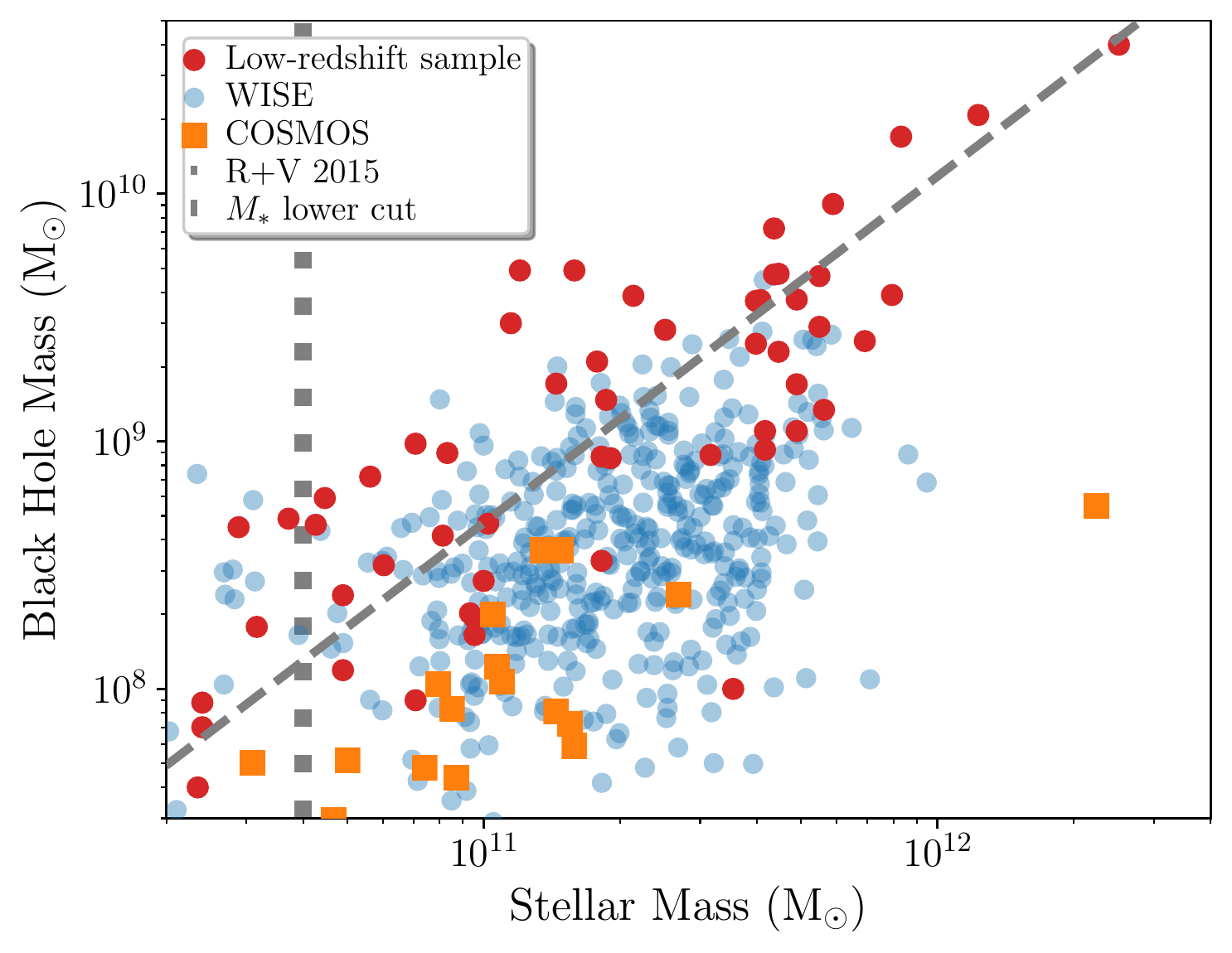}~~~
 \includegraphics[width=0.49\linewidth]{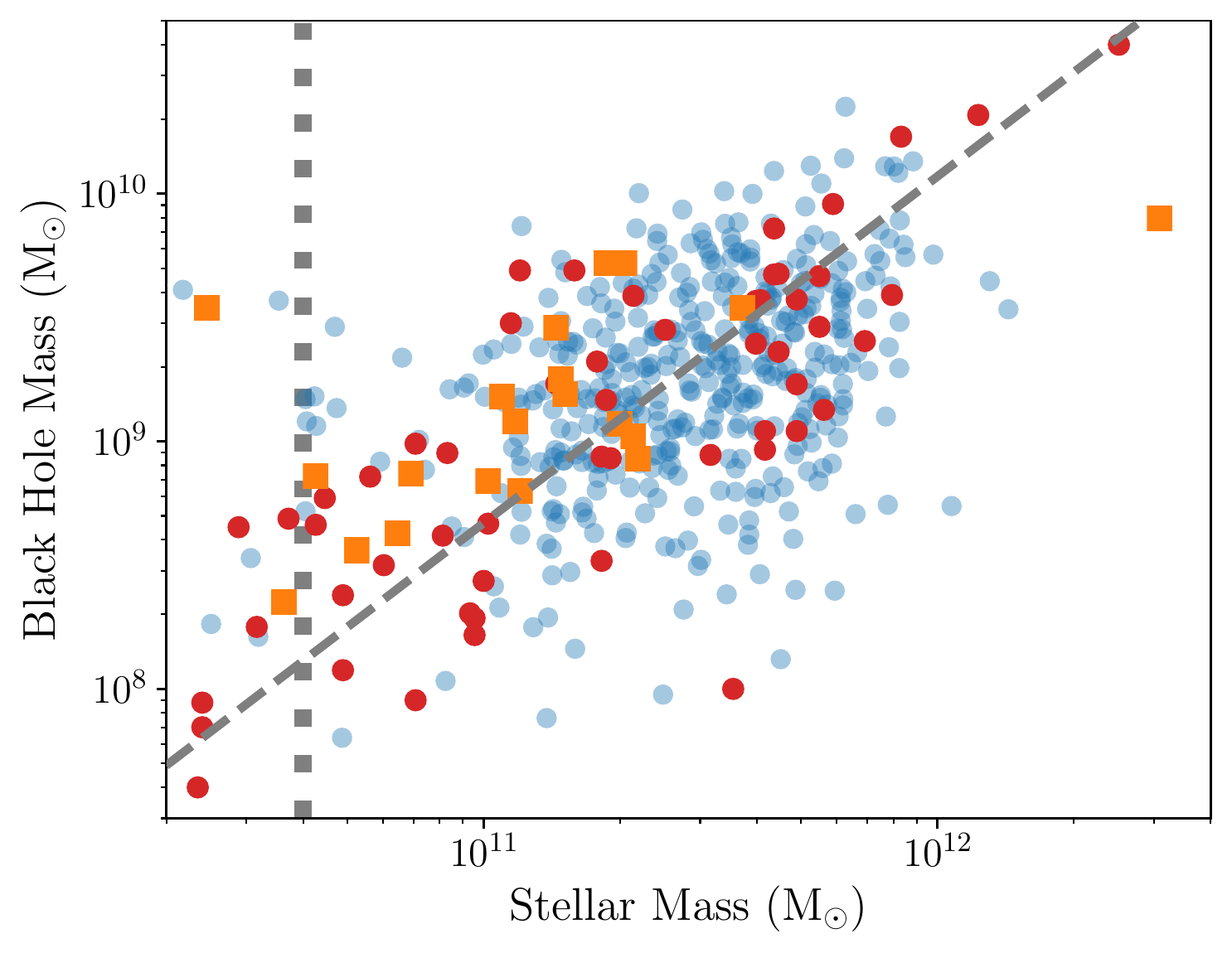} 
\caption{{\itshape Left:} The high-redshift (Figures \ref{fig:samplewise},\ref{fig:samplecosmos}) and quiescent local (Figure \ref{fig:samplelocal}) samples in the $M_{BH} - M_{*}$ plane as reported in the literature. The samples are offset from each other, consistent with the high redshift sample having lower SMBH masses for the same stellar mass. A similar offset is observed between local ellipticals and local AGN, though without a selection on the AGN host morphology \citep{rein15,shank19}. {\itshape Right:} The high-redshift and quiescent local samples in the $M_{BH} - M_{*}$ plane, with the median translational offsets in Equations~(\ref{eq:medtau_wise}) through (\ref{eq:medtaustar_cosmos}) applied (see also Figure \ref{fig:anomaly}). The translational offsets bring the samples into agreement.}
\label{fig:compwbias}
\end{center}
\end{figure*}

Figure \ref{fig:compwbias} shows the high and low redshift samples in the $M_{BH} - M_{*}$ plane, before and after the median $\tau_{*}$ and $\tau_{BH}$ offsets are applied, showing that they bring the samples into agreement. The left panel of this figure, and the offsets we find, are consistent with \citet[][figure 8]{rein15}, except that our samples differ in redshift by $\Delta z\sim0.85$, and are restricted to elliptical galaxies.

These translational offsets may arise from physical changes in stellar and SMBH mass, as well as selection and measurement bias. To determine what physical changes in stellar and SMBH mass are compatible with our results, we construct a model for the selection and measurement biases our samples are subject to. These biases are:

\begin{itemize}

\item Stellar mass bias ($B_{m,M_*}$): Differences in stellar mass arising from different measurement methods, different assumptions on the stellar IMF or M/L ratio, and passive stellar evolution between the two epochs.

\item Accretion bias ($B_g$): a bias in SMBH masses in AGN, arising since they have not completed their current accretion phase, during which they may gain further mass.

\item Dynamical selection bias ($B_{s,dyn}$): a bias in low redshift SMBH masses determined from stellar dynamical modeling, arising from the need to resolve the SMBH sphere of influence \citep{shank16a}.

\item Virial measurement bias ($B_{m,vir}$): a bias in SMBH masses in AGN measured via the single-epoch virial approach.

\item Virial selection bias ($B_{s,vir}$): a bias in SMBH masses in AGN, arising because the accretion luminosities of more massive SMBHs will, on average, be higher.

\end{itemize}

\begin{deluxetable}{rlc}
\tablehead{
\colhead{Parameter} & \colhead{Range (dex)} & \colhead{Discussion}
}
\tablecaption{Bias contributions to bias-corrected translations $\Delta\tau_{BH}$ and $\Delta\tau_*$ of the high redshift samples, relative to the low redshift sample.
  Ranges denote uniform probability.
\label{tbl:priors}}
\startdata
$B_{m,M_*}$               & $[0.08, 0.33]$  & \S\ref{sec:prior_B*} \\
$B_g$                     & $[0, 0.1]$     & \S\ref{sec:prior_Bg} \\
$B_{s,dyn}$               & $[0, 0.3]$     & \S\ref{sec:prior_Bdyn} \\
$B_{s,vir}$               & $[-0.3, -0.1]$ & \S\ref{sec:prior_Bvir} \\
$B_{m,vir}$ (\ion{Mg}{2}) & $[-0.4, 0.1]$ & \S\ref{sec:prior_Bvir}\\
$B_{m,vir}$ (H$\beta$).   & $[-0.2, 0.1]$ & \S\ref{sec:prior_Bvir}
\enddata
\end{deluxetable}

We summarize our adopted distributions for each of these biases in Table \ref{tbl:priors}. A detailed description of each bias is given in Appendix \ref{sec:descbias}.

The measured values of $\tau_{*}$ from the WISE and COSMOS samples are consistent with the expected prior $B_{m,M_*}$ arising from measurement bias alone. The measured value of $\tau_{BH}$ from the WISE sample is however in substantial tension with the expected offset from selection and measurement bias. The sum of the medians of the expected SMBH biases is $-0.15$\,dex, compared to $\tau_{BH}=0.70$\,dex, a difference of $0.85$\,dex. The measured $\tau_{BH}$ from the COSMOS sample is in even greater tension: $-0.15$\,dex vs. $\tau_{BH}=1.15$\,dex.

To explore this further, we assemble a combined distribution that includes all the SMBH expected biases from Table \ref{tbl:priors},

\begin{align}\label{collectivebh}
  B_{BH} := B_g + B_{s,dyn} + B_{s,vir} + B_{m,vir}.
\end{align}

\noindent We then define two new parameters

\begin{align}\label{eq:medtau}
\Delta \tau_{BH} & := \tau_{BH} - B_{BH} \\
\Delta \tau_*    & := \tau_{*} - B_{m,M_*}
\end{align}

\noindent Thus, $\Delta \tau_{BH}$ and $\Delta \tau_{*}$ parameterize the translational offsets in the SMBH and stellar mass axes, after selection and measurement bias have been accounted for.

\begin{figure*}
\includegraphics[width=\linewidth]{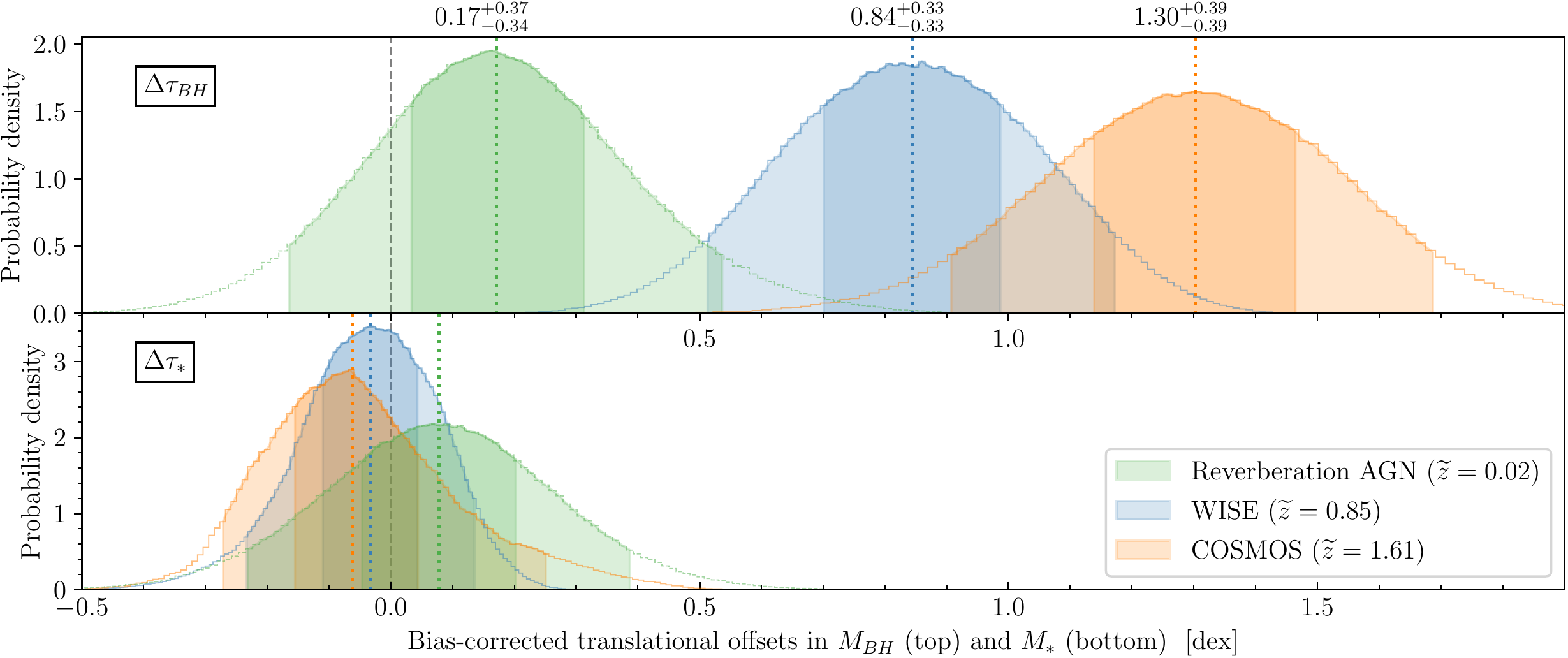}
\caption{\label{fig:tension} Posterior distributions for $\Delta\tau_{BH}$ (top) and $\Delta\tau_*$ (bottom), the bias-corrected translational offsets within the $M_{BH} - M_*$ plane, between the local quiescent sample and the WISE, COSMOS, and RM AGN samples. Shaded regions indicate $50\%$ (dark) and $90\%$ (light) confidence. If all the samples passively evolve, we expect $\Delta\tau_{BH} = \Delta\tau_* = 0$. All samples are consistent with $\Delta\tau_* = 0$. The only sample consistent with $\Delta\tau_{BH} = 0$ is the RM AGN. The median $\Delta\tau_{BH}$ for WISE and COSMOS signify, respectively, $\sim 7\times$ and $\sim 20\times$ growth of SMBH mass between their epochs.
}
\end{figure*}

In Figure~\ref{fig:tension}, we display marginalized posterior distributions for $\Delta \tau_{BH}$ and $\Delta \tau_{*}$. The recovered values for WISE are:

\begin{align}
\Delta \tau_{BH} & = +0.84~^{+0.33}_{-0.33}~\mathrm{dex}, \\
\Delta \tau_{*} & = -0.03~^{+0.17}_{-0.20}~\mathrm{dex},
\end{align}

\noindent The recovered values for COSMOS are:

\begin{align}
\Delta \tau_{BH} & = +1.30~^{+0.39}_{-0.40}~\mathrm{dex} \\
\Delta \tau_{*}  & = -0.06~^{+0.31}_{-0.21}~\mathrm{dex},
\end{align}

\noindent The recovered values for the RM AGN sample are:

\begin{align}
\Delta \tau_{BH} & = +0.17~^{+0.36}_{-0.34}~\mathrm{dex} \\
\Delta \tau_{*}  & = +0.08~^{+0.31}_{-0.31}~\mathrm{dex},
\end{align}

\noindent all with uncertainties at 90\% confidence.

In the appendix, we explore the sensitivity of our result to choices made in the analysis. In \S\ref{sec:hizch} we test against the choice of the high redshift sample. In \S\ref{stelbh} we explore the effects of varying the $M_{*}$ and $M_{BH}$ selections. In \S\ref{sec:likelihood} we explore the impact of changing both the test within the likelihood function, and the likelihood function itself. In \S\ref{sec:injections} we perform injection tests using artificial sources built from the WISE sample to assess the accuracy of our analysis.

Finally, to facilitate comparison with previous work, we compute the redshift evolution in $M_{BH}/M_*$ inferred from the WISE and local quiescent samples. The $\Delta\tau_{BH} = 0.84$\,dex is relative to biases that already assume an average $+0.05$\,dex growth in $M_{BH}$ due to remaining accretion during the optical AGN phase. As the WISE sample are observed before they have completed this growth, the redshift evolution in $M_{BH}/M_*$ is computed from the sum of these offsets: $0.89$\,dex. The result is equivalent to negative evolution in $M_{BH}/M_*$ with redshift:

\begin{align}
\frac{M_{BH}}{M_{*}} = \left(1 + z\right)^{3.5\pm1.4}
\end{align}

\noindent at $90\%$ confidence.

\section{Discussion}\label{sec:discussion}
We find that the SMBHs in massive, red-sequence elliptical galaxies have grown in mass relative to the stellar mass by a factor of seven from $z\sim1$ to $z=0$, and a factor of twenty from $z\sim2$ to $z=0$. We first compare our result to expectations from cosmological simulations in \S\ref{compsim}. We then explore possible reasons for our result in \S\ref{reasonfirst} - \S\ref{reasonlast}. We defer discussion of one further interpretation, cosmologically coupled astrophysical black holes \citep[e.g.][]{farja07,guariento2012realistic,maciel2015timedep,cro19,cro20,cro21,croker2022well}, to Farrah et al (2022, in preparation).

\subsection{Comparison to Cosmological Simulations}\label{compsim}
Cosmological simulations of galaxy assembly mostly predict insignificant change in the $M_{BH} - M_{*}$ relation among massive galaxies at $z<1$ \citep{habouzit21,habouzit22,zhangsim22}. Between $z\sim1$ and $z=0$ the change in $M_{BH}$ relative to $M_{*}$ in $M_{*}\gtrsim4\times10^{10}$\,M$_{\odot}$ systems is $-0.1$\,dex (Illustris, Horizon AGN, also e.g. \citealt{martin18}), $0$\,dex (TNG 100, TNG 300, EAGLE, TRINITY), and $+0.1$\,dex (SIMBA). Our result is in potential tension with these findings. These simulations, however, do not provide detailed morphological information, and do not study the evolution of the $M_{BH} - M_{*}$ relation in ellipticals alone,  making a matched selection to our study difficult.  Other simulations which find consistency with literature studies of the $M_{BH}/M_{*}$ ratio \citep{ni22sims,ding22} generally compare with AGN at all redshifts, and without regard to host type, making them distinct from our study.

\subsection{The expected biases are incorrect}\label{reasonfirst}
We now consider if the adopted bias distributions are incorrect. Assuming that there is no substantial error in $B_{m,M_*}$, this requires that $B_{BH}$ (Equation \ref{collectivebh}) is incorrect by $\sim 0.8 - 1.3$ dex. We thus consider each SMBH bias in Table \ref{tbl:priors} in turn.

Some virial selection bias is almost certainly present in the high redshift sample, so the most conservative value for $B_{s,vir}$ is $-0.1$\,dex \citep{treu07}. For virial measurement bias, the spread in reported disparities between RM and single-epoch masses is large, but most objects give larger single-epoch masses. It is reasonable to assume that the most conservative median value for $B_{m,vir}$ is $-0.1$\,dex.

This leaves $\sim0.6 - 1.1$\,dex of SMBH growth to account for. It is challenging to accommodate this as accretion in the remaining AGN phase. Doing so would require accretion at the maximum efficiency for a Kerr black hole (42\%, \citealt{bardeen1970}) at a substantial fraction of the Eddington limit for the duration of the final optical phase ($\sim 20$\,Myrs, c.f. Appendix~\ref{sec:prior_Bg}). This is in tension with the Eddington ratios in Figure \ref{fig:samplewisehists}, and with statistical constraints on accretion efficiency, though the assumptions behind this calculation are simplistic. This leaves dynamical selection bias. Explaining our result would require $B_{s,dyn} \simeq [0.6, 1.1]$ instead of our assumed $B_{s,dyn} = [0.0, 0.3]$, among SMBHs of mass $\gtrsim10^{8.5}$\,$M_{\odot}$. Based on Figure \ref{fig:samplelocal} and \citet{shank17}, this seems implausible.

A final issue for pure bias explanations is that the magnitude of the SMBH mass offset depends on redshift. None of the SMBH biases, with the possible exception of virial measurement bias, should significantly depend on redshift, and none of them have been shown to do so. We thus do not perceive a way in which biases alone can explain our result. This does however motivate a deeper understanding of selection and measurement bias in all SMBH mass measures, as an essential component in understanding the cosmic evolution of the SMBH mass function.

\subsection{The low redshift sample assembles at $z\gtrsim2$}
A way to explain our result is that the high redshift sample does not evolve into the low redshift sample, because the low redshift sample completed their assembly at $z\gtrsim2$, earlier than the WISE or COSMOS samples, {\itshape and} if the higher redshift assembly preferentially grows SMBHs, relative to $z\lesssim2$ channels.

We cannot exclude this possibility, but it seems unlikely. There is strong evidence from both observations \citep[e.g.][]{hast22} and simulations \citep{correa19} for the emergence of the red sequence at $z\lesssim1$. Moreover, stellar population ages in $z\lesssim0.1$ ellipticals \citep[e.g.][]{zhusdss10,forbes16,escudero18,sanro19,lacerna20,werle20,dolfi21,johnston22}, and in brightest cluster galaxies (BCGs) \citep{umanzor21}, are consistent with significant mass assembly at $z\sim1$. Numerical simulations also support this idea \citep{rodgo16,rosito19,habouzit21,habouzit21comp}. Finally, if our low-redshift sample did complete assembly at $z>2$ then we may expect closer agreement with the COSMOS, rather than the WISE, sample, but the opposite effect is observed. Furthermore, we do not see evidence for a `break' at $M_{*} \sim 2\times10^{11}$M$_{\odot}$ which might be expected in such a scenario \citep{kraj18a}.\footnote{A detailed investigation would need individual formation histories of our low-redshift sample, but this information is not available. Some of the sample have single-stellar population (SSP) ages \citep[e.g.][]{mcder15}, but these ages are challenging to interpret as they are sometimes older than the age of the Universe. Moreover, SSP ages can differ by factors of several for the same galaxy between different studies \citep[e.g.][]{georgiev12}.}

\subsection{The high redshift sample evolves into a subset of the low-redshift sample}
There are two ways in which the high redshift sample could evolve into a subset of the low-redshift sample:

\begin{enumerate}

\item The high redshift sample does not evolve further, and comprises the low $M_{BH}/M_{*}$ `tail' of the low redshift sample. The low redshift objects with higher $M_{BH}/M_{*}$ ratios arise from high redshift objects that grow significantly in stellar {\itshape and} SMBH mass since $z\sim0.8$.

\item The high redshift sample evolves into a low redshift population with, on average, higher SMBH {\itshape and} stellar masses than are seen in our low redshift sample. Our low redshift sample arises from objects at $z\gtrsim0.8$ that also grow significantly in both SMBH and stellar mass, starting (at $z\sim0.8$) with stellar masses below our stellar mass cut of $4\times10^{10}$M$_{\odot}$.

\end{enumerate}

The first of these scenarios encounters significant observational challenges. It would require that $\sim10^{10}$M$_{\odot}$ ellipticals at $z\sim0.8$ increase in stellar and SMBH mass by $\sim1$\,dex each by $z\sim0$. This would require 1-2 further mergers with a gas-rich progenitor, or 3-4 major `dry' mergers. This would imply that most ellipticals at $z<0.8$ should be merging, but this is not observed.

The second possibility seems plausible, as the co-moving volume of the universe over $0<z<0.16$ is about 45 times smaller than the volume of the universe over $0.7<z<0.9$. However, for this scenario to be true, most of the WISE sample would have to evolve into galaxies with $M_{*}\gtrsim10^{12}$M$_{\odot}$ by $z\lesssim0.2$. The required spatial density is about 11 sources per co-moving Gpc$^{3}$, implying that we should observe about ten such sources in our low-redshift sample. Yet, only 1-2 are observed. Moreover, recent studies suggest that the spatial density of very massive galaxies at high redshift may be underestimated \citep{gao21}.

A consistency check can be made by considering stellar mass to halo mass relations, and predicted halo merger trees with redshift. Observations suggest a halo mass of $\sim10^{12.5}$M$_{\odot}$ for a $M_{*}\sim10^{11}$M$_{\odot}$ galaxy at both $z=1$ and $z=0$, and with similar halo masses for galaxies across our mass range at both redshifts \citep{girelli20}. Cosmological simulations are consistent with this \citep{dubois21,Moster2012}. The morphological dependence of the evolution of the stellar to halo mass relation with redshift is not well constrained (though see e.g. \citealt{correa20}). Models also suggest that $\gtrsim10^{12}$M$_{\odot}$ halos undergo $\lesssim1$ major merger between $z=1$ and $z=0$ \citep{parkinson08,bose22}. We conclude that it is reasonable to assume our high redshift sample evolves mostly passively into our low redshift sample.

\subsection{The antecedents of the low-redshift sample are missed in the high redshift selection}
We next consider if the low redshift sample undergoes a final assembly episode over the same epoch as the WISE and COSMOS samples, but are not the same population. This could arise if the ancestors to the low redshift sample are missed in the high redshift samples because their hosts are too dim to detect, or if their SMBHs are accreting at an extremely low rate. Neither of these possibilities seems likely. Our imposed stellar mass cut is within the detection limit of both surveys, and we can think of no reason why more massive SMBHs in these samples would accrete more slowly on average. A final possibility is that the SED-based selection of the WISE sample does not also give a reliable morphology-based selection of early-type hosts. We do not believe this is likely, as our SED and $E(B-V)$ selections are strict, but we cannot test this possibility here.

\subsection{Evolution in a more fundamental scaling relation}
We may be observing evolution in a different, more fundamental, scaling relation (\citealt{denicola19}, see also \citealt{scott13,sahu19,sahu22}). For example, if the high and low redshift samples had systematically different velocity dispersions, $\sigma$, and the $M_{*} - \sigma$ relation evolved with redshift, then we could see this as evolution in the $M_{BH} - M_{*}$ relation. We cannot rule this out, but we do not know how such a selection bias could occur in our sample.

\subsection{Preferential conversion of molecular gas mass to SMBH mass}
The $M_{BH}/M_{*}$ ratio in $z\lesssim1$ ellipticals could evolve via preferential conversion of molecular gas to SMBH mass. This requires that the SMBHs accrete $\sim10^{9}$\,M$_{\odot}$ of molecular gas while the stellar mass increases by less than $\sim10^{10}$\,M$_{\odot}$.

Several mechanisms could in principle accomplish this, but all face challenges. A `major' merger with a gas-rich galaxy might suffice, if the encounter parameters strongly favored dissipation of angular momentum in the molecular gas so it could funnel to the nucleus. This is in some senses plausible,  it has recently been shown that gas-rich major mergers can lead to the central SMBH growing by $\sim1$\,dex, with the stellar mass increasing more modestly \citep{farrah22}. Such an event would, however, lead to rapid gas accretion in a relatively short burst, resulting in an AGN. It also seems unlikely that encounter parameters could globally favor the channeling of molecular gas to the nuclear regions in major mergers. 

Another possibility is one or more `minor' mergers with gas-rich dwarf galaxies. In general, minor mergers significantly influence the evolution of elliptical galaxies over $0.5\lesssim z \lesssim 1.0$ \citep[e.g.][]{bezanson09,naab09,hopkins10,kaviraj11} A few such mergers could in principle provide enough molecular gas to grow the SMBHs by $\lesssim10^{9}$M$_{\odot}$. However, minor mergers of ellipticals with gas-rich satellites have been observed to trigger star formation, meaning that at least some of the accreted gas is converted to stellar mass \citep{kaviraj11,kaviraj13,kaviraj14,rutkow14,jeong22,woodrum22}. Preferential growth of the SMBH through minor gas-rich mergers would also require encounter parameters that favor funneling of gas to the nucleus of the remnant. 

A non-merger based possibility is accretion of low-angular momentum gas from the intergalactic medium (IGM). This possibility would also result in an AGN. It would also only be possible in rich clusters, whereas our local sample span a full range of environments.

A final possibility is slow and steady accretion of molecular gas. At face value this is plausible since molecular tori exist around SMBHs in some galaxies \citep[e.g.][]{combes19}. It could also produce redshift dependence in the SMBH mass offset of the sign that we observe. However, this possibility also faces serious challenges. Local ellipticals have {\itshape total} molecular gas masses in the range $10^{8} - 10^{10}$M$_{\odot}$, as do massive, passively evolving galaxies with low SFRs at $z\sim0.7$ \citep{spil18}. This means that the SMBHs would have to accrete the majority of their hosts' gas content for this to be viable. Moreover, if this scenario were true, we may expect that the morphologies of the molecular gas in ellipticals to be concentrated towards the nuclear regions. However, molecular gas morphologies in local ellipticals are often smooth, symmetric, and with low Gini coefficients \citep{davis22}. Other studies find that the molecular gas traces the stellar light in many cases, on spatial scales of $\sim$kpc \citep{alatao13,davis13,sansom19}. Furthermore, the most actively star-forming local ellipticals tend to be the most gas-rich \citep{combes07}, which could suggest a stronger link between the molecular gas reservoir and star formation. Finally, the accretion rate required for this scenario is, on average, $\sim10^{-1}$\,M$_{\odot}$yr$^{-1}$. Such a high, average accretion rate in passively evolving galaxies seems unlikely given the accretion rates measured by the EHT.  For Sgr A$^{*}$ the accretion rate is $\lesssim10^{-7}$\,M$_{\odot}$yr$^{-1}$ \citep{sagaccrete} and for M87 it is $\sim10^{-3}$\,M$_{\odot}$yr$^{-1}$ \citep{m87accrete}.  Both are over an order of magnitude lower than the average required to align our high and low redshift samples.

\subsection{Dry mergers change the $M_{BH}/M_{*}$ ratio}
The $M_{BH}/M_{*}$ ratio could be changed between $z\sim1$ and $z\sim0$ by `dry' mergers in which the other galaxy has a relatively more massive SMBH. While such dry mergers are not expected to be common at $z<1$, especially at higher masses \citep[e.g.][]{bell05,lopez12,lee17}, a few such mergers are feasible. This possibility would however require the `other' galaxy to have more extreme ratios of $M_{BH}/M_{*}>0.1$. Such systems are rare within the \citet{barrows21} catalog, at any redshift. %% We cannot, however, rule out a quiescent population of such systems.

\subsection{Stellar mass loss}\label{reasonlast}
Our result could be explained if the high redshift sample lose an appreciable fraction of their stellar mass as they evolve from $z\sim1$ to $z\sim0$. This scenario is worth considering as it has been proposed that, under certain circumstances such as tidal stripping within clusters, galaxies can lose an appreciable fraction of their stellar mass as a function of decreasing redshift \citep[e.g.][]{kimm11,tollet17}. The stellar mass loss predicted by this process is however $20 - 30\%$, and will only occur in galaxies that infall into clusters over our adopted redshift range, so we do not believe it can, on its own, account for the effect we observe.

\section{Conclusions}
We have measured the translational offset between samples of ellipticals in the $M_{BH} - M_{*}$ plane over $z \lesssim 2$. Two samples at $0.8<z<0.9$ and $0.7 < z < 2.4$ feature AGN signposting the final assembly of their elliptical hosts. One sample at $z\sim0$ represents their evolutionary descendants in the local universe: quiescent ellipticals. An additional sample at $z\sim 0$ also represents their evolutionary descendants in the local universe, but feature a `flare up' AGN that allows RM cross-check measurement of SMBH mass.

We expect negligible translational shifts within the $M_{BH}-M_*$ plane between our low- and high-redshift samples. The low redshift samples are consistent with this expectation. The two high redshift samples, however, are offset along the $M_{BH}$ axis. These offsets are consistent with $\sim 7\times$ SMBH mass growth from $z \simeq 1$ and $\sim 20\times$ SMBH mass growth from $z \simeq 1.6$. Offsets this large, and unequal, cannot be easily accounted within expectations for selection and measurement bias.

Our result is robust to the key choices made in the analysis. We recover consistent translations using samples from different high redshift surveys, under variation in the assumed bias model, and under different choices in analysis methods.

We discuss distinct evolutionary pathways for the high and low redshift samples, unaccounted for selection biases, and SMBH accretion and/or merger pathways as explanations for the measured growth. None of these scenarios, individually, can plausibly explain our results. We conclude that selection biases are more complicated than previously appreciated, or there is an additional channel for SMBH mass growth.

\begin{acknowledgments}
We thank the referee for a valuable report. We thank the David C. and Marzia C. Schainker Family for their financial support of required computations. M.~Zevin is supported by NASA through the NASA Hubble Fellowship grant HST-HF2-51474.001-A awarded by the Space Telescope Science Institute, which is operated by the Association of Universities for Research in Astronomy, Inc., for NASA, under contract NAS5-26555. G.~Tarl\'e acknowledges support through DoE Award DE-SC009193. The National Radio Astronomy Observatory is a facility of the National Science Foundation operated under cooperative agreement by Associated Universities, Inc.
\end{acknowledgments}

\appendix

\startlongtable
\begin{deluxetable*}{lrrrrrccc}
\tablehead{
\colhead{Name} & \colhead{RA} & \colhead{Dec} & \colhead{$z$} & \colhead{$M_{BH}$} & \colhead{$M_{*}$} & \colhead{Env}\tablenotemark{a} & \colhead{AGN}\tablenotemark{b}
}
\tablecaption{The low-redshift elliptical samples used in our analysis.\label{tbl:lowz_sample}}
\startdata
Holmberg 15A & 10.46029  & -9.30313  & 0.0554 & $10.60^{+0.10}_{-0.08}$ & $12.40\pm 0.10$ & $2$ &  $0$ \\ 
IC 1459      & 44.29420  & -36.46222 & 0.0060 & $ 9.39^{+0.03}_{-0.08}$ & $11.60\pm 0.09$ & $1$ &  $1$ \\ 
Abell 1836   & 210.41908 & -11.60758 & 0.0363 & $ 9.59^{+0.05}_{-0.05}$ & $11.90\pm 0.12$ & $2$ &  $0$ \\ 
Abell 3565   & 204.16167 & -33.95833 & 0.0123 & $ 9.04^{+0.09}_{-0.09}$ & $11.69\pm 0.12$ & $2$ &  $1$ \\ 
NGC 0524 	 & 21.19883  & 9.53883   & 0.0080 & $ 8.94^{+0.02}_{-0.04}$ & $11.26\pm 0.09$ & $1$ &  $1$ \\ 
NGC 0584     & 22.83646  & -6.86806  & 0.0060 & $ 8.29^{+0.01}_{-0.01}$ & $10.98\pm 0.15$ & $1$ &  $0$ \\ 
NGC 0821 	 & 32.08808  & 10.99492  & 0.0058 & $ 8.22^{+0.25}_{-0.16}$ & $10.98\pm 0.09$ & $0$ &  $0$ \\ 
NGC 1216 	 & 6.82724   & -9.61281  & 0.0179 & $ 9.69^{+0.19}_{-0.13}$ & $11.20\pm 0.10$ & $1$ &  $0$ \\ 
NGC 1271 	 & 49.79700  & 41.35325  & 0.0199 & $ 9.48^{+0.20}_{-0.12}$ & $11.06\pm 0.07$ & $2$ &  $0$ \\ 
NGC 1277 	 & 49.96454  & 41.57353  & 0.0168 & $ 9.69^{+0.17}_{-0.12}$ & $11.08\pm 0.07$ & $2$ &  $0$ \\ 
NGC 1332 	 & 51.57188  & -21.33522 & 0.0054 & $ 9.17^{+0.06}_{-0.06}$ & $11.27\pm 0.09$ & $1$ &  $0$ \\ 
NGC 1374 	 & 53.81912  & -35.22625 & 0.0043 & $ 8.77^{+0.04}_{-0.04}$ & $10.65\pm 0.09$ & $2$ &  $0$ \\ 
NGC 1399 	 & 54.62094  & -35.45066 & 0.0047 & $ 8.94^{+0.31}_{-0.31}$ & $11.50\pm 0.09$ & $2$ &  $1$ \\ 
NGC 1407 	 & 55.04942  & -18.58011 & 0.0059 & $ 9.67^{+0.04}_{-0.06}$ & $11.74\pm 0.09$ & $1$ &  $0$ \\ 
NGC 1453     & 56.61354  & -3.96878  & 0.0130 & $ 9.46^{+0.06}_{-0.06}$ & $11.74\pm 0.15$ & $1$ &  $0$ \\ 
NGC 1550 	 & 64.90802  & 2.40946   & 0.0124 & $ 9.59^{+0.09}_{-0.06}$ & $11.33\pm 0.09$ & $1$ &  $0$ \\ 
NGC 1600 	 & 67.91642  & -5.08625  & 0.0156 & $10.23^{+0.04}_{-0.04}$ & $11.92\pm 0.09$ & $0$ &  $0$ \\ 
NGC 2549 	 & 124.74313 & 57.80305  & 0.0035 & $ 7.18^{+0.70}_{-0.05}$ & $ 9.71\pm 0.09$ & $1$ &  $0$ \\ 
NGC 2693     & 134.24696 & 51.34744  & 0.0161 & $ 9.23^{+0.12}_{-0.09}$ & $11.69\pm 0.15$ & $1$ &  $1$ \\ 
NGC 2784 	 & 138.08125 & -24.17261 & 0.0023 & $ 8.23^{+0.03}_{-0.02}$ & $10.23\pm 0.15$ & $1$ &  $0$ \\ 
NGC 3091 	 & 150.05954 & -19.63696 & 0.0132 & $ 9.57^{+0.06}_{-0.01}$ & $11.61\pm 0.09$ & $1$ &  $0$ \\ 
NGC 3115 	 & 151.30825 & -7.71858  & 0.0022 & $ 8.95^{+0.16}_{-0.03}$ & $10.92\pm 0.09$ & $1$ &  $0$ \\ 
NGC 3245 	 & 156.82662 & 28.50744  & 0.0044 & $ 8.38^{+0.17}_{-0.05}$ & $10.69\pm 0.09$ & $1$ &  $1$ \\ 
NGC 3377 	 & 161.92638 & 13.98592  & 0.0022 & $ 8.25^{+0.32}_{-0.18}$ & $10.50\pm 0.09$ & $1$ &  $0$ \\ 
NGC 3379 	 & 161.95662 & 12.58162  & 0.0030 & $ 8.62^{+0.12}_{-0.10}$ & $10.91\pm 0.09$ & $1$ &  $0$ \\ 
NGC 3585 	 & 68.32121  & -26.75484 & 0.0048 & $ 8.52^{+0.08}_{-0.16}$ & $11.26\pm 0.09$ & $0$ &  $0$ \\ 
NGC 3608 	 & 161.95662 & 12.58162  & 0.0030 & $ 8.67^{+0.10}_{-0.08}$ & $11.01\pm 0.09$ & $1$ &  $0$ \\ 
NGC 3842 	 & 176.00896 & 19.94981  & 0.0208 & $ 9.96^{+0.16}_{-0.10}$ & $11.77\pm 0.09$ & $2$ &  $0$ \\ 
NGC 3923 	 & 177.75706 & -28.80602 & 0.0058 & $ 9.45^{+0.13}_{-0.13}$ & $11.40\pm 0.12$ & $1$ &  $0$ \\ 
NGC 4291 	 & 185.07583 & 75.37083  & 0.0059 & $ 8.99^{+0.16}_{-0.12}$ & $10.85\pm 0.09$ & $1$ &  $0$ \\ 
NGC 4339 	 & 185.89562 & 6.08175.  & 0.0042 & $ 7.60^{+0.30}_{-0.35}$ & $10.37\pm 0.12$ & $2$ &  $0$ \\ 
NGC 4342 	 & 185.91250 & 7.05400   & 0.0025 & $ 8.65^{+0.18}_{-0.18}$ & $10.46\pm 0.12$ & $1$ &  $0$ \\ 
NGC 4350 	 & 85.99111  & 16.69341  & 0.0040 & $ 8.86^{+0.60}_{-0.42}$ & $10.75\pm 0.12$ & $1$ &  $0$ \\ 
NGC 4374 	 & 186.26560 & 12.88698  & 0.0034 & $ 8.97^{+0.04}_{-0.04}$ & $11.62\pm 0.09$ & $2$ &  $1$ \\ 
NGC 4434 	 & 186.90285 & 8.15434   & 0.0036 & $ 7.85^{+0.15}_{-0.15}$ & $10.38\pm 0.12$ & $1$ &  $0$ \\ 
NGC 4472 	 & 187.44484 & 8.00048   & 0.0033 & $ 9.40^{+0.02}_{-0.09}$ & $11.84\pm 0.09$ & $2$ &  $0$ \\ 
NGC 4473 	 & 187.45363 & 13.42936  & 0.0075 & $ 7.95^{+0.30}_{-0.18}$ & $10.85\pm 0.09$ & $1$ &  $0$ \\ 
NGC 4486A    & 187.74046 & 12.27036  & 0.0025 & $ 7.18^{+0.22}_{-0.12}$ & $10.04\pm 0.09$ & $2$ &  $0$ \\ 
NGC 4486B    & 210.41908 & -11.60758 & 0.0363 & $ 8.76^{+0.24}_{-0.24}$ & $ 9.66\pm 0.12$ & $2$ &  $0$ \\ 
NGC 4526     & 188.51286 & 7.69952   & 0.0021 & $ 8.65^{+0.11}_{-0.12}$ & $11.02\pm 0.30$ & $2$ &  $0$ \\ 
NGC 4564     & 189.11243 & 11.43928  & 0.0038 & $ 7.94^{+0.14}_{-0.11}$ & $10.38\pm 0.09$ & $2$ &  $0$ \\ 
NGC 4570 	 & 189.22250 & 7.24664   & 0.0060 & $ 8.08^{+0.03}_{-0.03}$ & $10.69\pm 0.20$ & $2$ &  $0$ \\ 
NGC 4621     & 190.50935 & 11.64703  & 0.0016 & $ 8.60^{+0.08}_{-0.08}$ & $11.12\pm 0.12$ & $2$ &  $0$ \\ 
NGC 4649     & 190.91656 & 11.55271  & 0.0037 & $ 9.67^{+0.11}_{-0.09}$ & $11.64\pm 0.09$ & $2$ &  $0$ \\ 
NGC 4697     & 192.14949 & -5.80074  & 0.0041 & $ 8.31^{+0.12}_{-0.10}$ & $10.97\pm 0.09$ & $0$ &  $0$ \\ 
NGC 4742     & 192.95017 & -10.45472 & 0.0042 & $ 7.30^{+0.30}_{-0.18}$ & $10.35\pm 0.12$ & $1$ &  $0$ \\ 
NGC 4751     & 193.21162 & -42.65992 & 0.0070 & $ 9.23^{+0.05}_{-0.02}$ & $11.16\pm 0.09$ & $1$ &  $0$ \\ 
NGC 4889     & 195.03387 & 27.97700  & 0.0215 & $10.32^{+0.63}_{-0.25}$ & $12.09\pm 0.09$ & $2$ &  $0$ \\ 
NGC 5018     & 198.25430 & -19.51819 & 0.0094 & $ 8.00^{+0.10}_{-0.08}$ & $11.55\pm 0.12$ & $1$ &  $1$ \\ 
NGC 5077     & 199.88196 & -12.65696 & 0.0094 & $ 8.93^{+0.32}_{-0.18}$ & $11.28\pm 0.09$ & $1$ &  $0$ \\ 
NGC 5328     & 208.22214 & -28.48940 & 0.0158 & $ 9.68^{+0.23}_{-0.07}$ & $11.65\pm 0.09$ & $1$ &  $0$ \\ 
NGC 5419     & 210.91138 & -33.97826 & 0.0138 & $ 9.86^{+0.14}_{-0.14}$ & $11.64\pm 0.12$ & $2$ &  $0$ \\ 
NGC 5516     & 213.97788 & -48.11486 & 0.0138 & $ 9.57^{+0.14}_{-0.01}$ & $11.60\pm 0.09$ & $1$ &  $0$ \\ 
NGC 5576     & 215.26533 & 3.27100   & 0.0050 & $ 8.44^{+0.15}_{-0.10}$ & $11.00\pm 0.09$ & $1$ &  $0$ \\ 
NGC 5845     & 226.50338 & 1.63380   & 0.0056 & $ 8.69^{+0.16}_{-0.12}$ & $10.57\pm 0.09$ & $1$ &  $0$ \\ 
NGC 5846     & 226.62202 & 1.60563   & 0.0057 & $ 9.04^{+0.06}_{-0.06}$ & $11.62\pm 0.16$ & $2$ &  $1$ \\ 
NGC 6086     & 243.14805 & 29.48479  & 0.0313 & $ 9.57^{+0.16}_{-0.17}$ & $11.69\pm 0.09$ & $2$ &  $0$ \\ 
NGC 6861     & 301.83117 & -48.37022 & 0.0094 & $ 9.32^{+0.02}_{-0.11}$ & $11.25\pm 0.09$ & $1$ &  $0$ \\ 
NGC 6958 	 & 312.17746 & -37.99742 & 0.0091 & $ 8.66^{+0.17}_{-0.12}$ & $10.63\pm 0.15$ & $2$ &  $0$ \\ 
NGC 7049 	 & 319.75126 & -48.56216 & 0.0076 & $ 8.50^{+0.13}_{-0.10}$ & $10.78\pm 0.15$ & $1$ &  $0$ \\ 
NGC 7457     & 345.24973 & 30.14494  & 0.0028 & $ 6.95^{+0.18}_{-0.19}$ & $ 9.56\pm 0.09$ & $1$ &  $0$ \\ 
NGC 7619     & 350.06055 & 8.20625   & 0.0126 & $ 9.36^{+0.02}_{-0.18}$ & $11.65\pm 0.09$ & $2$ &  $0$ \\ 
NGC 7768     & 357.74408 & 27.14739  & 0.0267 & $ 9.13^{+0.16}_{-0.14}$ & $11.75\pm 0.09$ & $2$ &  $0$ \\ 
Messier 87   & 187.70593 & 12.39112  & 0.0043 & $ 9.81^{+0.01}_{-0.01}$ & $11.72\pm 0.30$ & $2$ &  $2$ \\ 
Mrk 6        & 103.05105 & 74.42707  & 0.0195 & $ 8.10^{+0.04}_{-0.04}$ & $10.82\pm 0.30$ & $1$ &  $2$ \\ 
Mrk 509      & 311.04058 & -10.72348 & 0.0344 & $ 8.05^{+0.04}_{-0.03}$ & $10.40\pm 0.30$ & --- &  $2$ \\ 
Mrk 1501     & 2.62919.  & 10.97486  & 0.0872 & $ 8.07^{+0.17}_{-0.12}$ & $10.49\pm 0.30$ & --- &  $2$ \\ 
3C 390.3     & 280.53746 & 79.77142  & 0.0561 & $ 8.64^{+0.05}_{-0.04}$ & $10.66\pm 0.30$ & --- &  $2$ \\ 
PG 1307085   & 197.44584 & 8.33007   & 0.1538 & $ 8.54^{+0.16}_{-0.09}$ & $10.55\pm 0.30$ & --- &  $2$ \\ 
PG 1411442   & 213.45138 & 44.00388  & 0.0896 & $ 8.54^{+0.17}_{-0.12}$ & $10.69\pm 0.30$ & --- &  $2$ \\ 
Mrk 1506     & 68.29623  & 5.35434   & 0.0330 & $ 7.75^{+0.05}_{-0.04}$ & $10.70\pm 0.30$ & --- &  $2$ \\ 
Mrk 1095     & 79.04759  & -0.14983  & 0.0327 & $ 8.07^{+0.06}_{-0.05}$ & $10.68\pm 0.30$ & --- &  $2$ \\ 
PG 1617+175  & 245.04704 & 17.40769  & 0.1146 & $ 8.67^{+0.13}_{-0.08}$ & $ 9.74\pm 0.30$ & --- &  $2$ \\ 
3C 273       & 187.27792 & 2.05239   & 0.1583 & $ 8.84^{+0.11}_{-0.08}$ & $10.37\pm 0.30$ & --- &  $2$ \\ 
PG 0923+201  & 141.47798 & 19.90143  & 0.1927 & $ 9.07^{+0.06}_{-0.04}$ & $11.08\pm 0.40$ & --- &  $2$ \\ 
Mrk 876      & 243.48825 & 65.71933  & 0.1211 & $ 8.34^{+0.27}_{-0.16}$ & $11.34\pm 0.30$ & --- &  $2$ \\ 
PG 1700+518  & 255.35333 & 51.82222  & 0.2890 & $ 8.79^{+0.10}_{-0.09}$ & $10.69\pm 0.30$ & --- &  $2$ \\ 
\enddata
\tablenotetext{a}{Environment flag. $0$: field, $1$: group or pair, $2$: cluster.}
\tablenotetext{b}{Activity flag. $0$: quiescent, $1$: weak radio jet or hard X-ray source, $2$: Reverberation-mapped or EHT measured AGN.}
\end{deluxetable*}

\section{Analysis Pipeline Details}\label{sec:pipeline}
The \texttt{DYNESTY} sampler can operate in many different configurations that affect run-time performance and measurement fidelity. All analyses were performed with the simple nested sampler, to allow straightforward checkpointing and parallel operation. We adopted the \texttt{multi} bounding method, the \texttt{rwalk} exploration method, and 3000-6000 live points. We regarded analyses as converged when the change in accumulated evidence per iteration dropped below $\Delta\ln z = 0.01$.

The high and low redshift catalogs do not provide posterior draws, so we interpret reported uncertainties as $1 \sigma$ asymmetric (dimidated) Gaussian errors \citep{Barlow:2003xcj}. Posterior draws $\mathbf{d}_h$ and $\mathbf{d}_\ell$ from both high and low redshift catalogs, respectively, are then drawn from normalized asymmetric distributions in $M_*$ and $M_{BH}$, constructed per object.

Because the bias $B_{BH}$ is a sum of four other bias distributions, straightforward draws from all these can be slow. For uniform distributions $\left\{[\alpha_1, \beta_1], \dots, [\alpha_N, \beta_N]\right\}$, \citet{kamgar1995distribution} provide a convenient expression for the distribution of their sum,

\begin{align}
  B_{BH}(y) &\propto (-1)^N \sum_{\sigma \in \Sigma} \left(\prod_{s_i \in \sigma}^N s_i \right) \chi(\sigma, y)^{N-1} \Theta\left[\chi(\sigma, y)\right]
\end{align}

\noindent where $\Sigma$ indexes over all possible combinations of $N$ signs, $\Theta$ is the Heaviside step, and

\begin{align}
  \chi(\sigma, y) &:= y - \frac{1}{2}\sum_{i=1}^N \left[(\beta_i + \alpha_i) + s_i(\beta_i - \alpha_i) \right].
\end{align}

\noindent This expression allows draws from a single distribution instead of several, thus accelerating parameter estimation.

Because there is no physical model for the distribution of elliptical galaxies in the $M_{BH} - M_{*}$ plane, we construct a likelihood function built from non-parametric tests. There is some flexibility in how to proceed, because the stellar mass selection functions of the high and low redshift samples are mismatched and impossible to align analytically. For this reason, a two-dimensional quasi-parametric distribution test \citep[e.g.][]{fasano1987multidimensional} may not be appropriate. We adopt a likelihood function from the product of probabilities computed with the 2-sample test of \citet{epps1986omnibus} (Epps), applied to the marginalized posterior data draws

\begin{align}
p\big(\mathbf{d}_h, \mathbf{d}_\ell | \bm{\tau}\big) &:= \prod_v \Epps\left[\left<S\left(\mathbf{d}_h + \bm{\tau}\right)\right>_{M_v}, \left<\mathbf{d}_\ell\right>_{M_v}\right].
\label{eqn:projection_likelihood}
\end{align}

\noindent Here, $v \in \left\{*, BH\right\}$ and $S$ denotes a selection function to guarantee sample completeness and red sequence massive ellipticals

\begin{align}
S[(M_{BH}, M_*)] := \begin{cases}
        (M_{BH}, M_*) & M_* \geqslant M_\mathrm{cut} \\
        \varnothing & \mathrm{otherwise}
\end{cases}.
\end{align}

\noindent Ellipticals below $\sim 10^{10}M_\odot$ may also form via a range of channels, including gradual gas accretion, rather than major mergers followed by quiescence \citep[e.g.][]{tilman22}.

We have chosen the Epps-Singleton test following \citet[][Table VII]{epps1986omnibus}, as it has the most statistical power given our sample sizes. The ``projection'' likelihood defined in Equation~(\ref{eqn:projection_likelihood}) is not the only possible approach to mismatched stellar mass selection functions. The performance of our analysis under other likelihood functions, and with the likelihood of Equation~(\ref{eqn:projection_likelihood}) under other 1D non-parameteric tests, can be found in Appendix~\ref{sec:likelihood}.

\section{Bias Expectations}\label{sec:descbias}
The local and high redshift samples differ in both selection, and stellar and SMBH mass measurement methods. Therefore, we expect these differences to bias the samples relative to each other in the $M_{BH}-M_*$ plane. We here describe these biases, and our assumptions for their distributions.

\subsection{Stellar mass biases}\label{sec:prior_B*}
The stellar masses of the high and low-redshift samples are computed in different ways; aperture photometry followed by SED modeling assuming a Chabrier IMF for the high redshift sample, and profile decomposition assuming a Kroupa IMF for the low-redshift sample. Consistently recomputing the stellar masses of the samples is not currently possible, due to the challenges in such approaches for very extended sources \citep{petty13}. However, a subset of our low-redshift sample are also presented in \citet{rein15}, who compute $M_{*}$ using the M/L ratio in \citet{zibetti09}, via a Chabrier IMF. This approach is closer in concept to that used in the high redshift sample, and allows us to approximately calibrate across the samples. On average, \citet{rein15} find stellar masses that are $0.33$\,dex lower than those in \citet{zhu21}. Solely converting from a Kroupa to a Chabrier IMF would introduce a difference of $0.08$\,dex. We conclude that a stellar mass measurement bias applicable to the high redshift sample bracketing these values is reasonable.

The stellar mass selections of the high and low redshift samples, however, are also different. This leads to a bias that cannot easily be accounted for in standard approaches. We detail the relevant issues in Appendix~\ref{sec:likelihood}. There also exists the possibility that passive stellar evolution may change the mass-to-light ratio between $z\sim0.8$ and $z=0$, in ways that are hard to quantify given the uncertain stellar ages of the high redshift sample. For these reasons, we group stellar measurement and selection bias into $-0.5 \leqslant B_{m,M_*} \leqslant 0.5~\rm{dex}$.

\subsection{Early accretion growth}\label{sec:prior_Bg}
Assuming that the high redshift sample are chosen at random during the optical AGN phase, then their SMBH masses may increase by about half the total mass accreted during the optical AGN phase. Such accretion is not part of mass assembly in the passively evolving phase, so it must be accounted for prior to analysis.

A single AGN episode can increase SMBH mass by up to about an order of magnitude, but the majority of this occurs during the obscured phase. The mass accreted during the subsequent optical phase is likely a small fraction of this total \citep{mart05,baner15}. Moreover, for optical broad-line AGN, accretion rates are usually less than 10\% of the Eddington ratio \citep{kelly13}.

The fractional increase in mass due to accretion during the optical quasar phase, assuming constant Eddington ratio $\epsilon$ and accretion efficiency $\chi$, can be estimated from:

\begin{align}
  \frac{\dot{M}_{BH}}{M_{BH}} = \left(3.2\times10^7~\mathrm{s}\cdot\mathrm{yr}^{-1}\right)\frac{4 \pi G m_{p}\epsilon}{\chi c \sigma_{T}}.\label{eqn:optical_accretion}
\end{align}

Here $m_{p}$ is the proton mass, $G$ is Newton's constant, $c$ is light-speed, and $\sigma_{T}$ is the Thomson scattering cross-section. We assume an accretion efficiency of $\sim0.16$ \citep{sol82}, an Eddington ratio of $0.1$, and an optical quasar lifetime of $10-20$\,Myr \citep{hopkins05}.  This range in lifetimes is reasonable given current observational constraints \citep[e.g.][]{khry21}.  Integrating Equation~(\ref{eqn:optical_accretion}) and evaluating for the fractional change in mass over the optical phase gives a mass increase of $\lesssim 5\%$. We therefore conclude that an upper limit on the fractional mass increase due to accretion in the remainder of the optical quasar phase is $25\%$, or $0.1$\,dex, yielding a range of $0.0 \leqslant B_g \leqslant 0.1~\rm{dex}$.

\subsection{Dynamical SMBH bias}\label{sec:prior_Bdyn}
A selection bias has been proposed to exist in stellar dynamical SMBH masses \citep{shank16a,shank17}. The origin of this bias is that the ellipticals targeted for stellar dynamical SMBH measurements have higher than average central stellar velocity dispersions, due to the observational requirement to resolve the SMBH sphere of influence. Evidence for this bias is also seen in SMBH-galaxy scaling relations \citep{shank20}, and in the co-evolution of star formation and SMBH accretion up to $z=3.5$ \citep{carrero20}. This bias may be as high as an order of magnitude at $M_{BH} < 10^{7}$M$_{\odot}$ but is smaller at higher SMBH masses. At $M_{BH} > 10^{8}$M$_{\odot}$ it is thought to be at most a factor of two, and potentially negligible at $M_{BH} > 10^{9}$M$_{\odot}$. Based on the comparison in Figure \ref{fig:samplelocal}, we anticipate a dynamical selection bias of $0.0 \leqslant B_{s,dyn} \leqslant 0.3~\rm{dex}$.

Assessing the level of measurement bias in stellar dynamical SMBH masses is more challenging. The few existing studies that compare stellar dynamical SMBH masses to other approaches in the same systems find reasonable consistency (e.g. \citealt{robc21,thater22}). It is also notable that in M87, the stellar dynamical mass \citep{gebh11} is closer to the Event Horizon Telescope (EHT) mass than the gas dynamical mass \citep{walsh13}. We therefore assume there is no significant systematic measurement bias in stellar dynamical SMBH masses.

\begin{figure*}
\begin{center}
\includegraphics[width=0.49\linewidth]{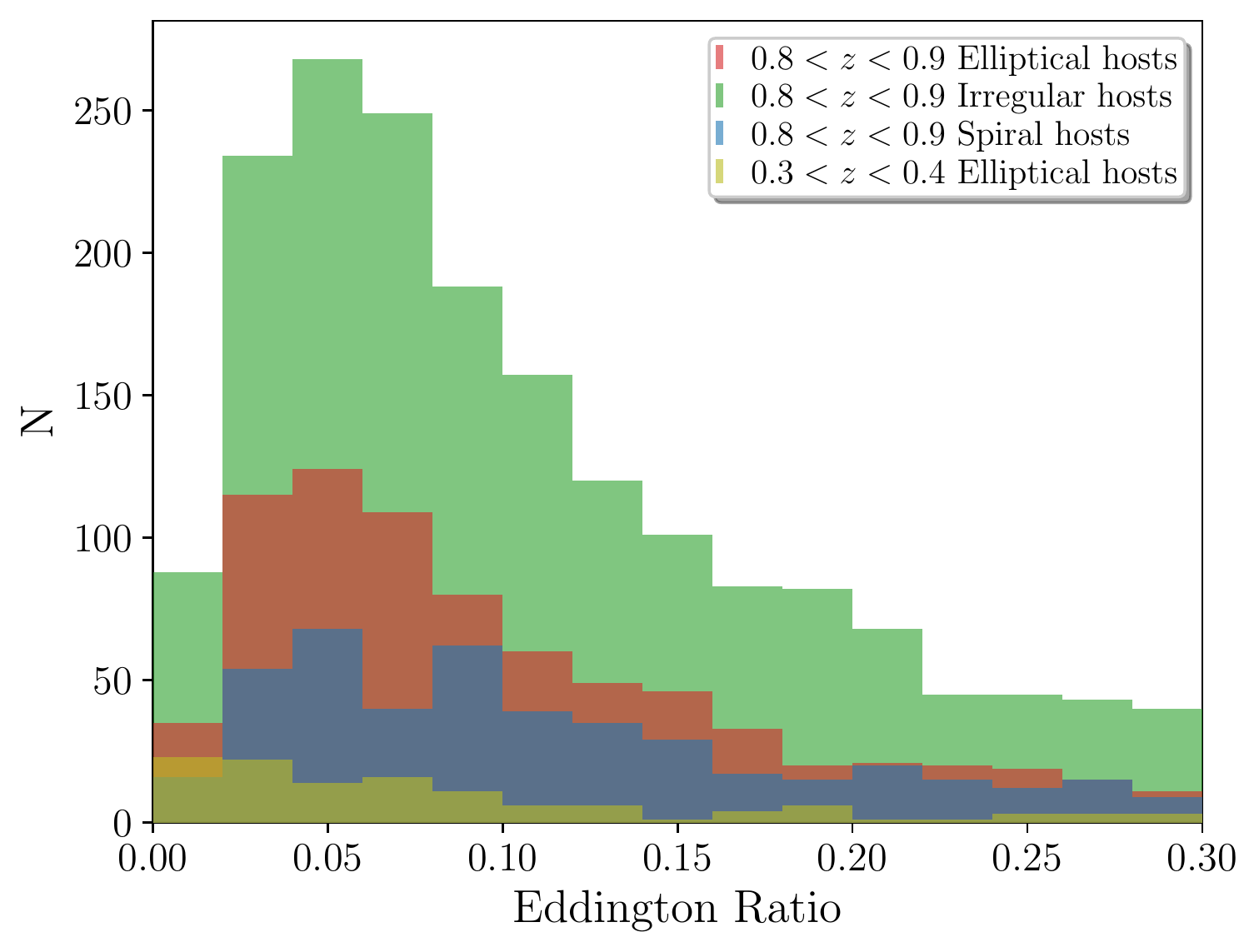}  \\
\caption{Histogram of the Eddington ratios of all AGN in \citet{barrows21} at $0.8<z<0.9$. The Eddington ratios are comparable across host SED types. Compared to AGN in elliptical hosts at $0.3<z<0.4$, we find little evolution in Eddington ratio across $\Delta z\simeq 0.5$ (\S\ref{sec:prior_Bvir}).
}
\label{fig:samplewisehists}
\end{center}
\end{figure*}

\subsection{Virial bias}\label{sec:prior_Bvir}

\subsubsection{Selection bias}
The high redshift sample is necessarily selected via the presence of AGN. This will bias the SMBH masses towards higher values, since AGN luminosity scales with SMBH mass \citep{richards06,ross13,runburg22}. A range of values have been proposed for this bias. \citet[][Figure~6]{lauer07} find that, depending on velocity dispersion, the bias lies between $0.0$ and $0.5$\,dex. \citet[][Figure~9]{treu07}, however, find that this bias is smaller, at $\sim 0.1$\,dex, but not zero. We therefore expect a virial bias of $-0.3 \leqslant B_{s,vir} \leqslant -0.1~\rm{dex}$, where the lower bound is the logspace average of the $0.1$ and $0.5$ literature bounds.

A related potential bias is redshift evolution in Eddington ratio. If Eddington ratios are higher at higher redshift then this may lead to smaller observed SMBH masses in flux-limited samples at higher redshift. We cannot comprehensively examine this possibility here, but we do not believe it is likely to be dominant. Figure \ref{fig:samplewisehists} shows that Eddington ratios do not vary significantly as a function of host SED type at $\widetilde{z} \simeq0.85$. Moreover, comparing to a sample of AGN in elliptical hosts at $0.3<z<0.4$ selected from the WISE catalog in the same way as the $0.8<z<0.9$ sample shows only small changes in Eddington ratio with redshift; $\epsilon=0.069^{+0.180}_{-0.050}$ at $0.3<z<0.4$ and $\epsilon=0.096^{+0.210}_{-0.057}$ at $0.8<z<0.9$ (68\% confidence). We therefore assume that this bias is subdominant to the virial selection bias described above. 

Finally, \citet{schwi11} have suggested a further bias in the $M_{BH} - M_{*}$ relation among AGN; the length of the AGN duty cycle biases the $M_{BH}/M_{*}$ ratio low if the duty cycle shortens as $M_{BH}$ increases, and vice versa. Since the duty cycles in our AGN sample are hard to constrain, we do not attempt to correct for this bias, but note it as a possible contaminant.

\subsubsection{Measurement bias}
As described in e.g. \citet{hokim14,yong16,shank19} and \citealt{lisha21}, single-epoch virial SMBH masses can be biased towards higher values, by a factor of 2-3 (see also e.g. \citealt{shen06,shenkelly12,mej17,wangs20,maithil22}).  Direct comparison between single-epoch virial and RM masses in the same AGN at $z>0.3$ are sparse and are usually based on data spanning less than $\sim5$ years, but allow for some basic comparisons. For \ion{Mg}{2}, comparing the RM masses in \citet{homay20} with the single-epoch masses in \citet{raks20} shows that the majority of objects have mass offsets in the range $-0.1 - 0.4$\,dex, though with some outside this range.   Instead, cross-matching the \citet{yu21rv} sample with \citet{raks20} yields one match (DES J021612.83-044634.10), for which the RM mass is 0.61\,dex lower than the single-epoch virial mass.   For H$\beta$, comparing the RM masses in \citet{bao22,lijen21} to the single-epoch masses reveals a similar range, though slightly closer to zero, with the majority of objects having a single-epoch mass that is higher by up to a factor of 2-3 than the RM mass, though some objects lie outside of this range, on either side.   Finally, for comparison, for \ion{C}{4} there are six objects published in \citet{lira18}, for which the single-epoch masses are higher by a factor of 2-10. We therefore adopt: $-0.4 \leqslant B_{m,vir} \leqslant 0.1~\mathrm{dex}$ for SMBH masses measured with \ion{Mg}{2}, and $-0.2 \leqslant B_{m,vir} \leqslant 0.1~\mathrm{dex}$ for SMBH masses measured with H$\beta$.

Other potential measurement biases exist. An example is the choice of virial factor. The SMBH masses we use are derived using $f_{vir}\sim4.5$, but \citet{shank19} argue that $f_{vir}\simeq3.3$ is more appropriate. This amounts to a downward correction to the SMBH masses of $\sim0.14$\,dex.  Another example is that it has been argued that using the line FWHM to measure single-epoch virial masses may overestimate the masses of high mass SMBHs \citep{dalla20}.  We do not include these as separate corrections, but assume that they are folded into the range of adopted virial measurement bias.

\subsubsection{Reverberation mapped AGN}\label{biases:revagn}
The RM AGN sample requires a modified set of biases. There is no virial measurement bias ($B_{m,vir} = 0$), and virial selection bias becomes $-0.1 \leqslant B_{s,vir} \leqslant 0.0$ as the sample are more in line with those of \citet{treu07} in terms of redshift and luminosity.

Finally, the use of RM, rather than single-epoch virial, BH masses excludes several samples: the WISE and SDSS samples at low redshift, the 244 broad-line AGN in \citet{rein15}, and the low luminosity AGN sample in \citet{busch14}. Moreover, in e.g. Figure~8 of \citet[][]{rein15}, and in \citet{shank19}, the local AGN (both RM and single-epoch virial) are offset below the quiescent early-type galaxies in the $M_{BH} - M_{*}$ plane, which seems at odds with our Figure~\ref{fig:samplelocal}. We hypothesize that this difference is due to the (mostly) non-elliptical host morphologies of the AGN in these samples, since late-type galaxies typically have smaller $M_{BH}/M_{*}$ ratios than early-type galaxies (e.g. \citealt[][Figure~12]{rein15}). Taking the sample in \citet{busch14}, restricting to early-type hosts with Sersic indices $n>3$, and accounting for differences in stellar M/L and IMF, leaves a sample of objects whose positions are consistent with our quiescent sample in the $M_{BH} - M_{*}$ plane. We note that the RM AGN in \citet{rein15} are all late-type, though host morphologies are not available for a representative number of their 244 broad-line AGN.

\section{Tests of stability}\label{sec:stabtests}

\subsection{Choice of high-redshift sample}\label{sec:hizch}
We here consider the impact on our results of using different high-redshift samples. First is a sample selected from \citet{barrows21}, but in the redshift range $0.7<z<0.8$. This is still within the formation epoch of the red sequence, but at a lower redshift than the main sample, and uses $H\beta$ rather than \ion{Mg}{2} to compute SMBH masses, with the SMBH masses again taken from \citet{raks20}. The prescription used to compute the H$\beta$ SMBH masses is identical to that used in the COSMOS sample.

Second are two samples drawn from the QSO hosts presented by \citet{licat21}. This sample overlaps the WISE sample in redshift, but is purely optically selected. Moreover, the stellar masses are not obtained by decomposing the AGN+host SED but instead by first subtracting the quasar point spread function (PSF) before measuring the host emission directly. While PSF subtraction from ground-based data is challenging, this sample thus gives a further check on the results from the WISE sample. Since the stellar masses in the SDSS sample are measured in PSF-subtracted optical imaging, there is the potential for bias towards luminous hosts, but we do not attempt to account for this. To select the quiescent hosts from this sample, we again band-merge with the \citep{raks20} catalog to obtain SMBH masses. We then impose the following selection criteria: a `quiescent' host galaxy type flag, and an SMBH mass with a `good' quality flag. The catalog also includes Sersic indices and radii. Since these quantities are hard to determine from ground-based, PSF-subtracted imaging, we use them only to perform a basic selection, demanding a Sersic index greater than unity. We then extract two sub-samples; one at $0.8<z<0.9$ with SMBH masses from \ion{Mg}{2}, and one at $0.7<z<0.8$ with SMBH masses from H$\beta$. The distribution of the samples in the $M_{BH} - M_{*}$ plane is similar to that of the WISE sample. The derived samples overlap negligibly with the WISE sample, so we treat them as independent.

\begin{figure}
\begin{center}
\includegraphics[width=0.48\columnwidth]{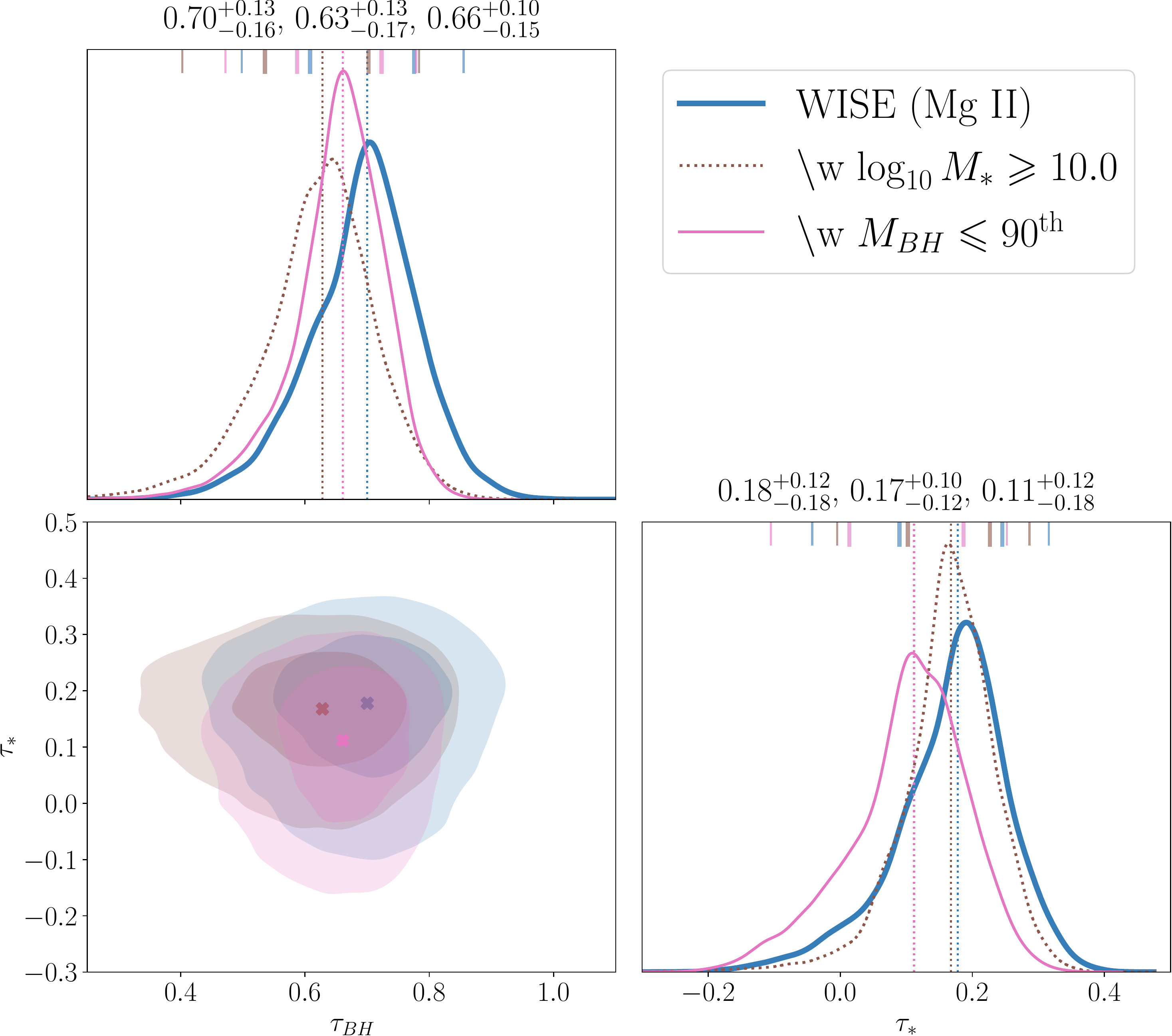} ~~
\includegraphics[width=0.48\columnwidth]{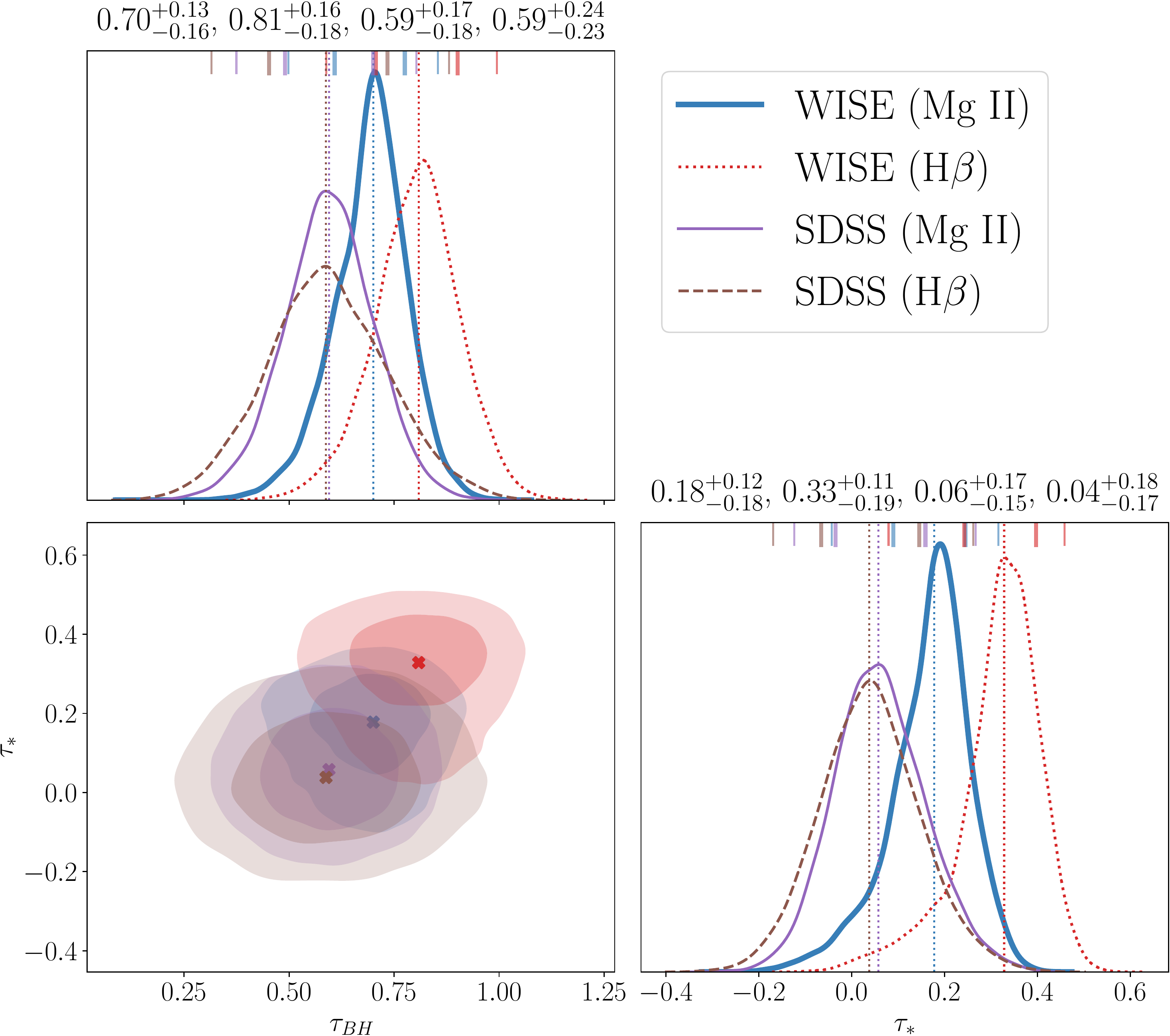}
\end{center}
\caption{\label{fig:allthecorners} {\itshape Top left:} The impact of adopting different subsets of the high and low redshift samples on derived posterior translations $\bm{\tau}$ within the $M_*-M_{BH}$ plane. Posterior distributions in the $\tau_{BH} - \tau_*$ plane are as follows: ``Main'' (blue, thick, solid) the main analysis; $\log_{10} M_* \geqslant 10.0$  changes the stellar mass cut to $1\times10^{10}$M$_{\odot}$; $M_{BH} \leqslant 90^{\mathrm{th}}$ excludes galaxies with SMBH masses in excess of their sample's top $10\%$. Marginalized distributions for $\tau_*$ and $\tau_{BH}$ are reported at $90\%$ confidence.
{\itshape Top right:} Comparison of the main WISE sample to the SDSS sample, showing impact of distinct spectral line on derived posterior translations $\bm{\tau}$ within the $M_*-M_{BH}$ plane. Marginalized distributions for $\tau_*$ and $\tau_{BH}$ are reported at $90\%$ confidence.}
\end{figure}

The analyses using the WISE and SDSS samples are compared in Figure \ref{fig:allthecorners}. The results are consistent across WISE and SDSS and different spectral lines.

\subsection{Stellar and SMBH mass selections}\label{stelbh}
The stellar masses of the high and low redshift samples are, by necessity, not computed consistently. Coupled with our analysis approach, this leads to two potential issues. First, the stellar masses of the high-redshift sample are computed with stellar population synthesis fitting. In most cases, this approach only allows for a limited set of star formation histories, which can affect the derived stellar masses. Second, the high-redshift sample will undergo passive stellar evolution from $z\sim0.8$ to $z\sim0$. Such passive evolution could lead to some of the offset we observe in the $M_{BH} - M_{*}$ plane.

Because our analysis approach determines consistency in $M_{BH} - M_{*}$ plane position between the high and low-redshift samples after a cut in stellar mass, rather than consistency in the $M_{BH}/M_{*}$ ratio, either of these possibilities could recast the effect we observe as a change in SMBH mass. In other words, from the perspective of diagonal alignment, it is conceivable that selecting $\log M_* \geqslant 10.6$ could introduce an artificial preference for translation in $M_{BH}$, instead of in $M_*$.

We investigate this possibility as follows. First, we note that the $\tau_{*}$ needed to diagonally align the high and low redshift samples if $\tau_{BH}=0$ is $-0.9$\,dex. In Figure~\ref{fig:allthecorners} (dotted), we restore objects with stellar mass $\geqslant 10^{10}~M_\odot$ in order to see whether the data prefer this route to alignment in the $M_{BH}-M_*$ plane. While the data do prefer a lower median $\tau_* = 0.07$\,dex, $95\%$ of the posterior distribution lies above $\tau_* = +0.05$\,dex when including lower-mass systems. We conclude that removing lower-mass systems has not preferentially biased our analysis to recovering large $\Delta\tau_{BH}$. In fact, in all analysis variants we have considered, the posterior for $\tau_*$ is almost entirely within $-0.2 \leqslant \tau_* \leqslant 0.6$, while still recovering an large $\Delta\tau_{BH}$. Passive stellar evolution is thus unlikely to have led to the large $\Delta\tau_{BH}$.

As an further check, we repeated the main analysis using the single $M_{BH}/M_{*}$ ratio, rather than both $M_{BH}$ and $M_*$, as diagnostics. In general, this approach degenerates the data and may lead to unphysical shifts in mass if restrictive priors are not asserted. Performing this analysis yields results consistent with these expectations; the stellar mass posterior reproduces the prior, and the uncertainties on the SMBH translation are in effect decided by the width of the stellar mass prior.

Finally, we consider the possibility that the most massive SMBHs in the low-redshift sample arise from a different channel to the rest of the low-redshift sample, and provide most of the measured translation. To test this, we exclude objects from the high and low redshift sample with SMBH mass within the top 10\% percentile of their sample. We find no significant effect on $\tau_{BH}$ or $\tau_*$. We conclude that our result is not driven by outlier systems.

\subsection{Non-parameteric and Likelihood Studies}
\label{sec:likelihood}
\begin{figure*}
\begin{center}
\includegraphics[width=0.48\linewidth]{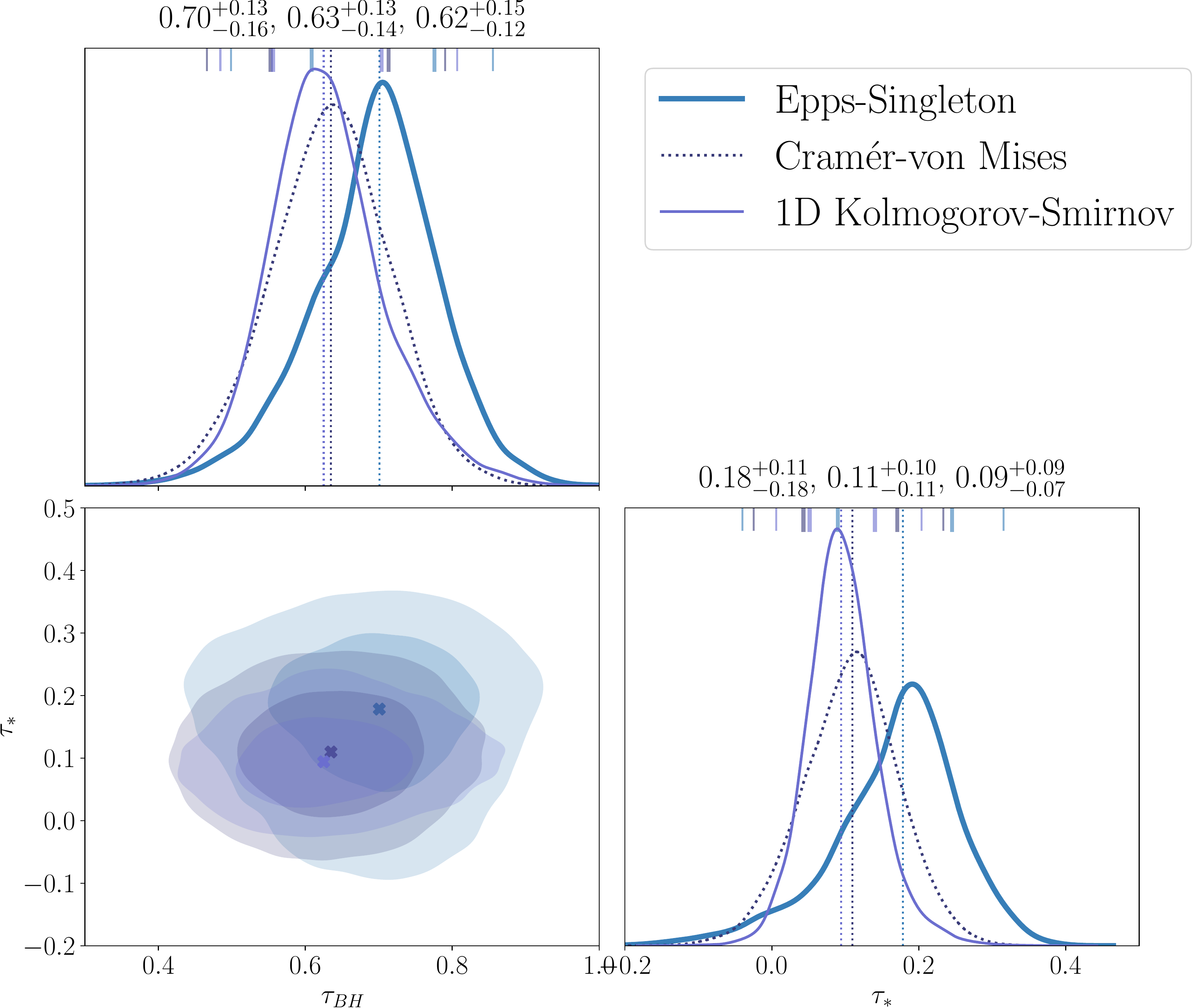}
\includegraphics[width=0.48\linewidth]{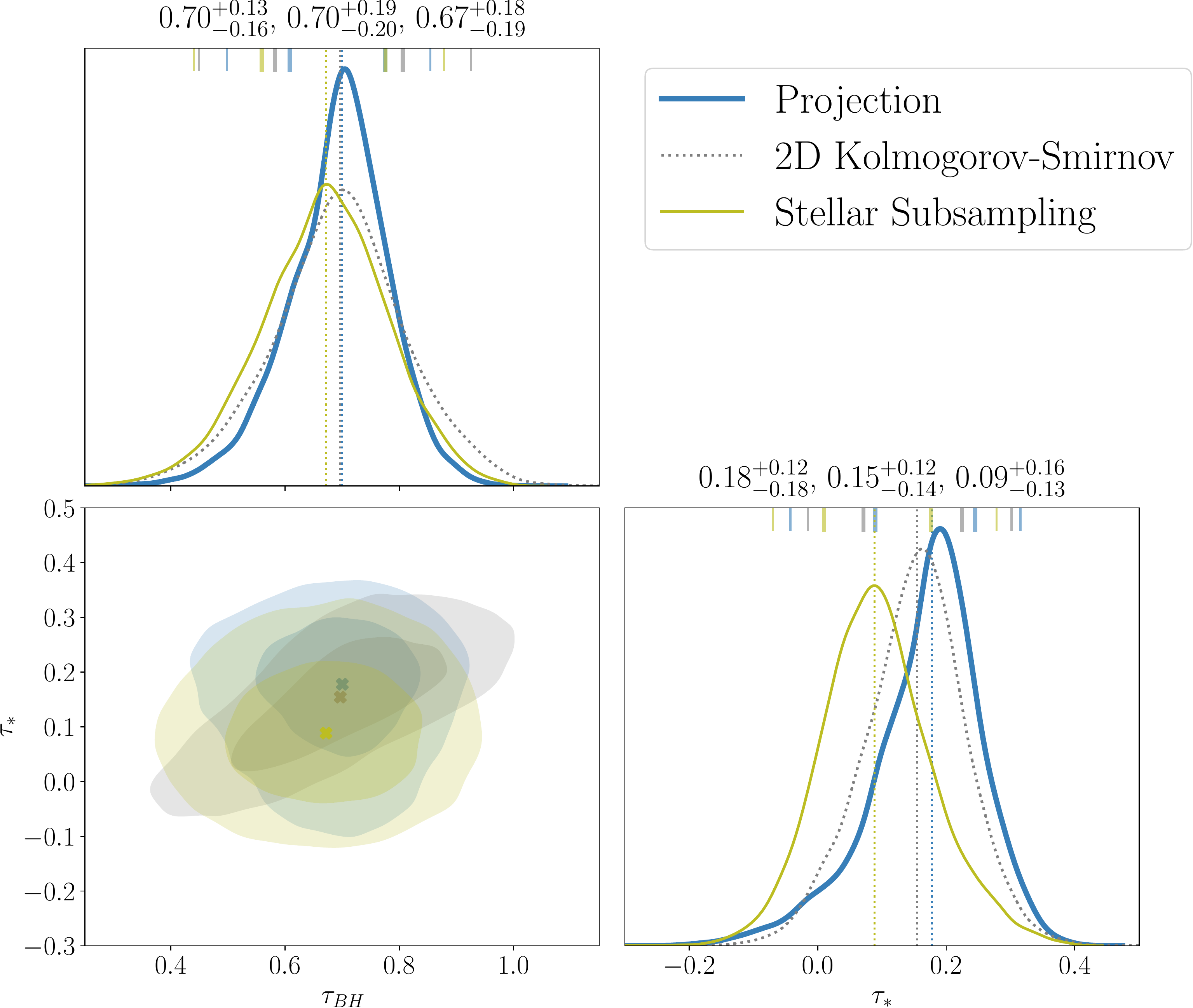}
\end{center}
\caption{\label{fig:stat_like_tests} (Left) Robustness to one-dimensional non-parametric statistical test within the projection likelihood adopted for the main analysis; (Right) Robustness to adopted likelihood function for the main analysis. Shown options represent different approaches to the unknown stellar mass bias selection function relative to high and low redshift samples.
  }
\end{figure*}

\begin{figure*}
\begin{center}
\includegraphics[width=0.49\linewidth]{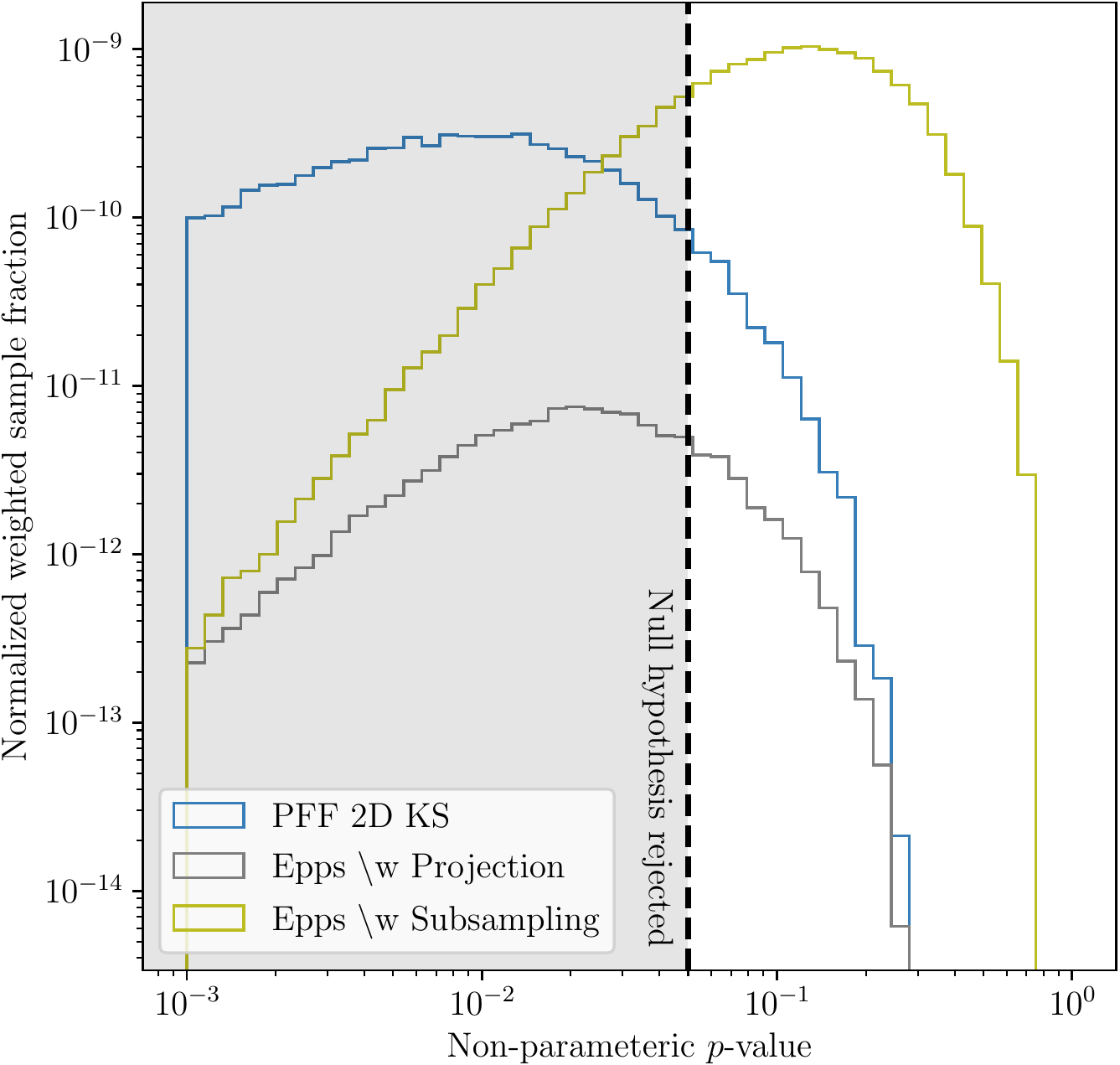}
  \end{center}
  \caption{\label{fig:pvalues} Non-parametric $p$-values of posterior $\tau_{BH} - \tau_*$ samples in the main analysis (WISE), under the different likelihood functions described in Figure~\ref{fig:stat_like_tests}. The 2D KS test returns a $p$-value directly, whereas projection and stellar subsampling return the product of $p$-values for distinct $\tau_{BH}$ and $\tau_*$ prior draws. Thus, for these latter two tests, the reported $p$-value is the geometric average. For all likelihood functions, there are regions of parameter space where the non-parametric test cannot reject the null hypothesis, i.e. the samples are plausibly ancestral. The 2D KS test performs worst, as expected due to stellar population mismatch. Subsampling successfully minimizes stellar selection mismatch bias: the bulk of evidence-weighted samples have large $p$-values. The illustrated fiducial value for rejection, $p \leqslant 0.05$ (dashed), is arbitrary and not used in any likelihood evaluation.
  }
\end{figure*}

We first consider the impact of choice of 1D non-parametric test within the likelihood  Equation~(\ref{eqn:projection_likelihood}). We compare against the lower-power, but faster to evaluate, Kolmogorov-Smirnov and Cram\'er-von Mises \citep{anderson1962distribution} tests. The results are displayed in Figure~\ref{fig:stat_like_tests} (left). All tests are statistically indistinguishable.

Stellar mass selection biases between the high and low redshift samples motivate consideration of alternate likelihood functions, which may be more or less sensitive to these biases. Neither the local or high redshift samples have straightforward selection functions in $M_{*}$. The local sample is heterogeneous, based in part on which objects are likely to yield a SMBH mass measure, and thus does not have a quantitative $M_{*}$ selection function. The high redshift samples are flux-limited, but the host light must be distinguished from the AGN light, meaning that the $M_{*}$ selection function is complicated.

To quantify the impact of the unknown $M_*$ selection function on $\Delta\tau_{BH}$, we have performed our main analysis with three distinct types of likelihood function:

\begin{enumerate}
\item{\emph{Naive}: A quasi-parametric 2D Kolmogorov-Smirnov test determines a likelihood in both $M_{BH}$ and $M_*$ simultaneously, despite the known mismatch.}
\item{\emph{Projected}: A projection likelihood fits the $M_{BH}$ and $M_*$ distributions separately via true non-parametric 1D tests, but weights their $p$-values equally. This decouples measurement of $\tau_{BH}$ from the anticipated degradation of fit due to stellar mass mismatch, but still allows recovery of $\tau_*$. This is the analysis adopted in the article text.}
\item{\emph{Subsampled}: Before applying the projected non-parametric tests, for each object in the low redshift realization, the five nearest neighbors in stellar mass are chosen, without repeats, from the high redshift population realization. In order to guarantee that the entire high redshift population is eventually probed, the nearest neighbors are taken from a random draw of half the size of the full high redshift sample. Note that this method does \emph{not} assume the ancestral hypothesis circularly: arbitrary evolution in $M_*$ can still occur, e.g. the nearest neighbors to a $5\times 10^9~M_\odot$ low redshift object might all have stellar masses $\sim 10^9~M_\odot$. This approach to minimizing the bias suffers from decreased statistics, as the effective high redshift sample is reduced by a factor of $\sim 4$, at least.
}
\end{enumerate}

The results of these three approaches are displayed in Figure~\ref{fig:stat_like_tests} (right), where they are statistically indistinguishable. With respect to recovery of $\tau_{BH}$, we conclude that stellar mismatch bias between the high and low redshift populations does not significantly contaminate the result.

The performances of these likelihoods are displayed in Figure~\ref{fig:pvalues}, where we show the evidence-weighted fraction of posterior draw non-parametric $p$-values, i.e. the probability that the two distributions were drawn from the same underlying distribution, after translation by $\bm{\tau}$. This visualization answers the question: of the draws used to estimate the posterior distribution in the $\tau_{BH}-\tau_*$ plane, how important were draws taken from regions where the ancestral hypothesis was plausible? Note that all likelihood methods probe regions of parameter space where the high and low redshift populations cannot be distinguished. The 2D KS test reflects overall low significance due to stellar mismatch, while the subsampling likelihood has effectively eliminated stellar mismatch. The projection method adopted in the main analysis is fast and compromises between both these approaches, increasing statistical power at the expense of overall likelihood due to stellar mismatch.

\subsection{Injection studies}\label{sec:injections}
\begin{figure}
\begin{center}
\includegraphics[width=0.48\linewidth]{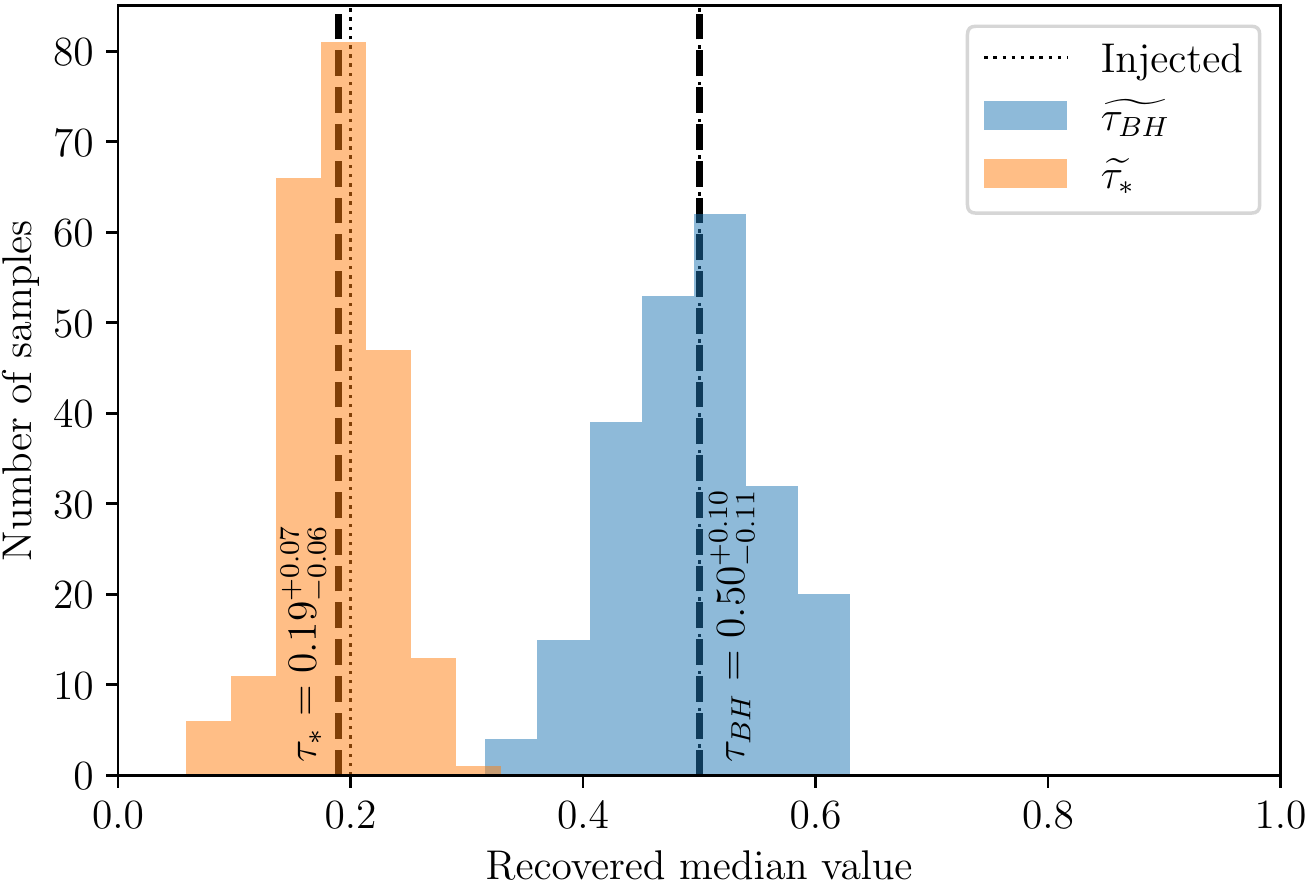}
\includegraphics[width=0.48\linewidth]{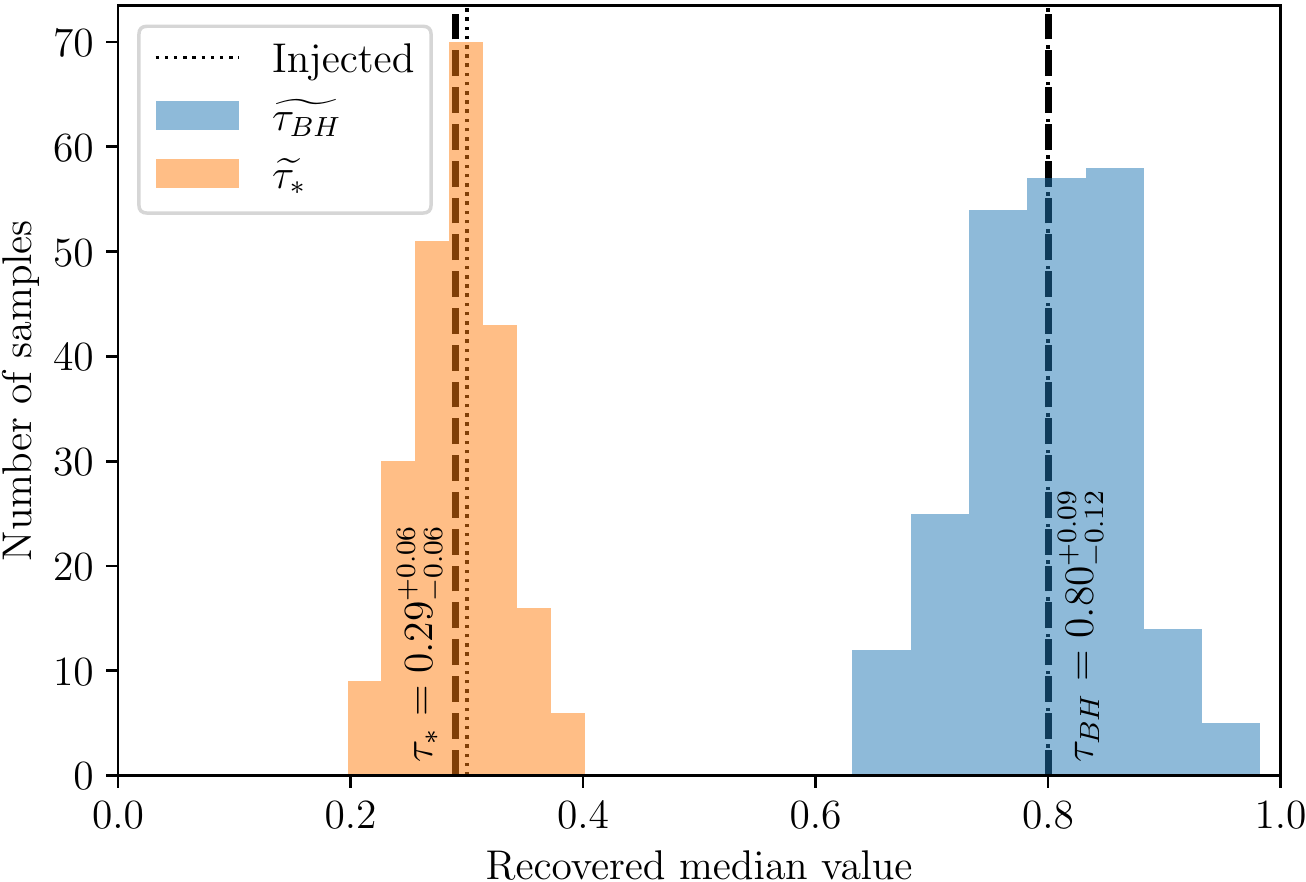}
\end{center}
\caption{\label{fig:injections} Performance of analysis pipeline in the recovery of injected $\tau_{BH}$ (blue) and $\tau_*$ (orange) from mock catalogs. Results show the distribution of 225 median values from the analysis of distinct mock catalog pairs. Injected values (dotted), median values (dashed), and 90\% confidences are displayed for each parameter. Recovery of the injected signal is clear and provides a measurement of the systematic uncertainty inherent to our analysis methodology.
  }
\end{figure}

In addition to measurement uncertainties on posterior data, there may be systematic uncertainty from analysis methodology. To gauge the ability of our analysis pipeline to recover $\tau_*$ and $\tau_{BH}$ from data, we proceed as follows. As our low-redshift samples are typically $\sim 50$ objects, we draw 50 objects from the high redshift sample and injected $\tau_* := \{0.2,0.3\}$ and $\tau_{BH} := \{0.5,0.8\}$ translations into this draw. The injected values for $\tau_{BH}$ were chosen to bracket analysis-variant measured values of $\tau_{BH}$ for the primary WISE Mg II sample. We then regarded the translated subsample as a mock low redshift sample, regarding the remaining $\sim 750$ distinct objects as a mock high redshift sample. Note that the mock low-redshift sample is constructed from the high-redshift sample so as to guarantee that both samples are drawn, without repeats, from an actual astrophysical distribution.

We then performed the \texttt{DYNESTY} analysis to $\Delta \ln z = 0.1$, sufficient to recover median $\tau_{BH}$ and $\tau_*$ to high precision. Posterior measurement uncertainties from the high redshift sample are included in the injection analysis within both mocks, even though the actual low redshift sample will have slightly different uncertainties, in order to estimate the ability of the pipeline to recover signal from actual data. This injection study was repeated $225$ times for distinct low and high redshift mock catalogs, producing the distribution of median values displayed in Figure~\ref{fig:injections}. From these studies, we find no systematic pull in $\tau_{BH}$. In $\tau_*$, we do find a systematic downward pull of $-0.01$~dex, independent of injected value.

\bibliography{ellipcc}{}
\bibliographystyle{aasjournal}

\end{document}